\documentclass[useAMS]{mn2e}
                                                                                
 \usepackage{times}

\usepackage{graphicx}
\def\>{$>$}
\def\<{$<$}

\def\simlt{\lower.5ex\hbox{$\; \buildrel < \over \sim \;$}}
\def\simgt{\lower.5ex\hbox{$\; \buildrel > \over \sim \;$}}
\def\ch2{$\chi^{2}$}

\def\ee{\'{e}}

\def\S{Section}

\def\sT{\sigma_{\rm T}}
\def\signuc{\sigma}

\def\be{\begin{equation}}
\def\ee{\end{equation}}
\def\beq{\begin{eqnarray}}
\def\eeq{\end{eqnarray}}

\def\RD{R_\Delta}
\def\tC{t_{\rm C}}
\def\TC{T_{\rm C}}

\def\A{{\cal A}}
\def\B{{\cal B}}
\def\C{{\cal C}}
\def\GA{\Gamma_{\cal A}}
\def\GB{\Gamma_{\cal B}}
\def\GC{\Gamma_{\cal C}}
\def\LA{L_{\cal A}}
\def\LB{L_{\cal B}}
\def\Lg{L_{\gamma}}
\def\Gn{\Gamma_n}
\def\Gcrit{\Gamma_{\rm crit}}
\def\Grel{\Gamma_{\rm rel}}

\def\Rdec{R_n}
\def\Rn{R_n}
\def\Rgg{R_{\gamma\gamma}}
\def\dn{\dot{n}}
\def\dnann{\dot{n}_{\rm ann}}
\def\texp{t_{\rm exp}}
\def\Ug{U_\gamma}
\def\dQnth{\dot{Q}_{\rm nth}}
  \def\dQep{\dot{Q}_{\rm th}}
  \def\Qep{Q_{\rm th}}
\def\fnu{f_\nu}
\def\fin{f_{\rm inel}}
\def\tT{\tau_{\rm T}}
\def\tp{\tau_p}
\def\tgg{\tau_{\gamma\gamma}}
\def\sgg{\sigma_{\gamma\gamma}}
\def\taunuc{\tau}

\def\dN{\dot{N}}
\def\dNann{\dot{N}_{\rm ann}}
\def\nann{n_{\rm ann}}

\def\Rph{R_\star}

\def\Eph{\bar{E}}
\def\Ep{E_{\rm peak}}
\def\dECoul{\dot{E}_{\rm Coul}}
\def\bprot{\beta_p}

\def\eff{\epsilon}

\def\M{{\cal M}}
\def\UKN{U_{\rm KN}}
\def\epKN{\bar{\varepsilon}}
\def\ath{a_{\rm th}}
\def\anth{a_{\rm nth}}
\def\tauR{\tau_0}
\def\effph{\eff_\star}

\def\ww{w}

\def\Lnth{L_{\rm nth}}
\def\kgg{\kappa_{\gamma\gamma}}
\def\ginj{\gamma_{\rm inj}}

 \newbox\grsign \setbox\grsign=\hbox{$>$} \newdimen\grdimen \grdimen=\ht\grsign
 \newbox\simlessbox \newbox\simgreatbox \newbox\simpropbox
 \setbox\simgreatbox=\hbox{\raise.5ex\hbox{$>$}\llap
      {\lower.5ex\hbox{$\sim$}}}\ht1=\grdimen\dp1=0pt
 \setbox\simlessbox=\hbox{\raise.5ex\hbox{$<$}\llap
      {\lower.5ex\hbox{$\sim$}}}\ht2=\grdimen\dp2=0pt
 \setbox\simpropbox=\hbox{\raise.5ex\hbox{$\propto$}\llap
      {\lower.5ex\hbox{$\sim$}}}\ht2=\grdimen\dp2=0pt
 \def\simgt{\mathrel{\copy\simgreatbox}}
 \def\simlt{\mathrel{\copy\simlessbox}}
 
\title[Collisional mechanism for GRB emission]
       {Collisional mechanism for GRB emission}
\author[Andrei M. Beloborodov]{Andrei M. Beloborodov\thanks{Also at 
Astro-Space Center of Lebedev 
Physical Institute, Profsojuznaja 84/32, Moscow 117810, Russia}\\
Physics Department and Columbia Astrophysics Laboratory, 
Columbia University, 538  West 120th Street New York, NY 10027;\\
amb@phys.columbia.edu}
\begin{document}
%
%
%
\maketitle
\label{firstpage}
\begin{abstract}
Nuclear and Coulomb collisions in GRB jets create a hot $e^\pm$ plasma.
This collisional heating starts when the jet is still opaque and extends 
to the transparent region. The $e^\pm$ plasma radiates its energy.
As a result, a large fraction of the jet energy is converted to escaping 
radiation with a well-defined spectrum. The process is simulated in detail
using the known rates of collisions and accurate calculations of radiative 
transfer in the expanding jet. The result reproduces the spectra of 
observed GRBs that typically peak near 1~MeV and extend to much higher 
energies with a photon index $\beta\sim -2.5$. This suggests that 
collisional heating may be the main mechanism for GRB emission.
\end{abstract}



\section{Introduction}

Cosmological gamma-ray bursts (GRBs) are associated with ultra-relativistic 
jets from short-lived powerful sources such as hyper-accreting, just-born 
black holes. The jet starts as an opaque blackbody fireball that accelerates, 
expands and releases its thermal radiation at the photospheric radius 
$R_\star$. One may expect a quasi-blackbody spectrum from such jets 
(Paczy\'nski 1986; Goodman 1986), similar to the relict radiation from the 
big bang. However, the simpe blackbody model is inconsistent with observations
(e.g. Preece et al. 2000). It is clear that some form of heating operates 
in the jet and changes its radiation from blackbody.
Heating may occur at radii $r<R_\star$ and change the photospheric radiation 
via Comptonization.
It may also occur at radii $r>R_\star$ and generate nonthermal 
synchrotron emission.

Two heating mechanisms are usually considered in GRB jets: internal shocks 
and dissipation of magnetic energy. 
The details of both mechanisms 
are uncertain as they depend on complicated {\it collisionless}
processes in the plasma.
A long-standing problem is the radiative efficiency of these processes. 

Recent observations by {\it Fermi} telescope provided new data
in a broad spectral range from 8 keV to $\sim 100$~GeV.
The data confirm the previous BATSE result that the prompt GRB spectrum 
typically peaks near MeV (Preece et al. 2000).
The typical spectrum is approximately described by the Band function 
(a smoothly broken power law), which extends to high-energy bands
with a photon index $\beta\sim -2.5$. The prompt GRB radiation is highly 
variable on timescales as short as millisecond, suggesting a small radius 
of emission, possibly comparable to the photospheric radius $\Rph$.\footnote{
Recent suggestions that the prompt $\gamma$-ray emission must come from a 
large radius $r\gg \Rph$ (e.g. Racusin et al. 2008; Abdo et al. 2009a) are 
based on incorrect assumptions (see \S~6.1).}

The spectrum of GRB emission at very high energies $E\simgt 1$~GeV remains 
so far uncertain. Multi-GeV photons overlapping 
the prompt MeV radiation have been detected in about 5 per cent of GRBs, 
however they may be produced by a distinct source at large radii,
e.g. by the blast wave from the explosion. 
The distinct source is visible in several GeV-emitting bursts
(e.g. Abdo et al. 2009b; Ryde et al. 2010) and likely present in all of them, 
obscuring the behavior of the prompt Band 
spectrum at high energies (cf. the debate over
GRB~080916C: Abdo et al. 2009a; Kumar \& Barniol Duran 2009; 
Ghisellini et al. 2009).

The standard theory
(e.g. Paczy\'nski 1990) predicts that jets with $\Gamma\sim 10^3$ must 
produce bright photospheric emission. Observational search for this emission 
usually assumed that it has a quasi-thermal spectrum 
(e.g. Ryde 2005). It was argued that the absence of the photospheric 
emission component would imply that the jet is magnetically dominated and 
cold, with negligible initial thermal energy (e.g. Daigne \& Mochkovitch 2002;
Zhang \& Pe'er 2009).

The main finding of the present paper is that the Band-type spectrum naturally 
forms in GRB jets as a result of {\it collisional} heating. 
Most of radiation produced by this mechanism is emitted near the photosphere 
$\Rph$ and therefore it may be called photospheric to a first approximation.
Our result supports the view that photospheric emission is not a rare 
quasi-thermal component; instead, it is the main Band component of GRB 
emission that is routinely observed in all bursts.

This paper considers the standard model of a baryonic jet 
with comparable numbers of neutrons and protons (our model would not 
work for jets that are completely dominated by magnetic field,  
with negligible baryonic loading).
  Before becoming transparent to radiation, the jet evolves to 
  the two-fluid or `compound' state:
a plasma with bulk Lorentz factor $\Gamma$ embeds a neutron flow 
with Lorentz factor $\Gamma_n<\Gamma$
(Derishev, Kocharovsky \& Kocharovsky 1999a; Bahcall \& M\'eszaros 2000; 
Fuller, Pruet \& Abazajian 2000; M\'esz\'aros \& Rees 2000;
Rossi, Beloborodov \& Rees 2006; Koers \& Giannios 2007). 
Regardless the details of their formation, compound jets 
with $\Gamma_n\ll \Gamma$ have a robust feature: 
nuclear collisions between the neutron and proton fluids
continually create multiple 
$e^\pm$ with energies $\sim m_\pi c^2\approx 140$~MeV.
  Their energy is immediately converted to 
  radiation before $e^\pm$ join the thermalized plasma.
  Nuclear collisions also heat the proton component of the jet, and protons 
  gradually drain their energy into thermalized $e^\pm$ plasma via Coulomb 
  collisions. 

Similarly to internal shocks, collisional heating taps
the kinetic energy of internal motions in the jet --- the streaming 
of plasma through the neutron component with a relative Lorentz factor 
$\Grel=\frac{1}{2}(\Gamma/\Gn+\Gn/\Gamma)$. 
In contrast to internal shocks, the heating is not confined 
to a shock front. It operates in volume. 
 
Several works previously proposed that some sort of volume heating 
shapes the spectrum of GRB emission (e.g. Thompson 1994; 
Ghisellini \& Celotti 1999; Stern \& Poutanen 2004; Rees \& M\'esz\'aros 
2005; Pe'er, M\'esz\'aros \& Rees 2005; Giannios \& Spruit 2007;
Ioka et al. 2007; Asano \& Terasawa 2009).
The models assumed some form of collisionless dissipation, which is 
difficult to calculate from first principles. In this context, two special 
features of our model should be noted:
                                                                                
\medskip

\noindent
(i) The collisional heating is robust, and its history in the expanding 
jet is well defined. The rate of collisions determines the radial 
dependence of the heating rate $\dot{Q}\propto r^{-2}$.
                                                           
\medskip

\noindent
(ii) The collisional heating injects energy into $e^\pm$ 
via two branches with comparable heating rates:

(a) Nuclear collisions maintain a continual $e^\pm$ cascade in the jet.

(b) Coulomb collisions in the two-temperature plasma\footnote{
     The thermalized $e^\pm$ are Compton-cooled and kept at a temperature
     much lower than the proton temperature (see \S~4.4).} 
continually transfer energy from protons to thermalized $e^\pm$. 

\medskip

\noindent
Branch (a) is important because it loads the jet 
with a large number of $e^\pm$ pairs and determines the photospheric
radius of the burst. On the other hand, it will be 
shown that branch (b) plays an important role in the formation of the GRB 
spectrum. The radiation emerging from a collisionally heated 
jet has a well defined spectrum, which can be calculated numerically. 
This radiative transfer
problem is solved in this paper using a Monte-Carlo code that tracks the 
evolution of photons and $e^\pm$ in the heated and expanding plasma flow.

The paper is organized as follows.
\S~2 gives a compact summary of the simplest 
model of GRB jets with no internal dissipation.
Such jets passively cool down as they expand, and eventually emit thermal 
radiation whose spectrum cuts off exponentially at $\sim$~MeV. 
It is inconsistent with the observed GRB spectra.
We use the model of a passively cooling jet as a benchmark and 
a first test problem for our radiative transfer code.

\S~3 describes neutron-loaded jets and formation of compound flows with 
$\Gn<\Gamma$. \S~4 describes the collisional radiative mechanism operating 
in compound flows. \S~5 presents the radiation spectrum received by distant 
observers.
The results are discussed in \S~6. 
\S~6 also discusses the possibility of additional emission 
that may be generated by neutron decay.

GRB outflows are believed to be beamed and therefore called `jets' throughout 
this paper. However, the results apply equally well to spherically symmetric 
outflows. As long as the opening angle of the explosion exceeds $1/\Gamma$, 
the jet near the axis is causally disconnected from its edge, 
and its dynamics is the same as that of a spherically symmetric flow.


\section{Thermal emission from passively cooling jets}

We focus in this paper on jets that are accelerated by thermal (radiation) 
pressure, with a subdominant magnetic field. This standard 
model is briefly summarized below (see e.g. Paczy\'nski 1990).

At small radii $r$, the GRB jet is in thermodynamic equilibrium and its 
luminosity is carried mainly by radiation 
$L\approx (4/3)caT^4\Gamma^2 4\pi r^2$
(hereafter we use the isotropic equivalent of luminosity, which would be 
produced by a spherically symmetric outflow of the same density and 
temperature).
As the jet expands adiabatically, the ratio of photon and baryon number 
densities $n_\gamma/n$ remains constant, i.e. effectively the photon number 
is conserved (similar to the cosmological big bang). 
The jet accelerates until the 
radiation energy density $\Ug=aT^4$ decreases below the rest-mass density 
$nm_pc^2$. Then its Lorentz factor saturates at the asymptotic $\Gamma$.
We will denote the characteristic saturation radius by $R_s$. 
For a radially expanding jet,
\be
\label{eq:R_s}
  R_s\approx \Gamma\, r_0,
\ee
where $r_0$ is the radius at the base of the jet, at the beginning of its 
acceleration. The photon-to-baryon ratio in the jet is given by
\be
\label{eq:ratio}
   \frac{n_\gamma}{n}\approx 240\,\Gamma\,r_{0,7}^{1/2}L_{52}^{-1/4},
\ee
and the jet energy {\it per photon} is 
\be
\label{eq:Eph0}
  \Eph_0\approx\frac{\Gamma m_pc^2 n}{n_\gamma}
      \approx 4\, r_{0,7}^{-1/2}L_{52}^{1/4} {\rm ~MeV}.
\ee

If no heat is generated by any dissipative processes, 
the thermal radiation trapped in the opaque flow continues to cool 
adiabatically at $r>R_s$ until it is released at the photosphere $\Rph$.
Between $R_s$ and $\Rph$, the radiation temperature decreases 
as $n^{\hat{\gamma}-1}$ where $\hat{\gamma}=4/3$ is the adiabatic index 
of radiation, which gives $\Eph\propto r^{-2/3}$.
The plasma has a small heat capacity (for $n\ll n_\gamma$) and simply 
tracks the radiation temperature. Electrons are thermally coupled to radiation
via Compton scattering, and ions maintain thermal equilibrium with electrons 
via Coulomb collisions at a common low temperature $kT\simlt 1$~keV in 
the rest frame of the jet.

The optical depth of the jet is given by
\be
\label{eq:tau_e}
  \tT=\frac{n\sT r}{\Gamma}=\frac{L\sT}{4\pi rm_pc^3\Gamma^3}
   \approx r_{10}^{-1}\,L_{52}\,\Gamma_3^{-3},
\ee
where $\sT=6.65\times 10^{-25}$~cm$^2$ is Thomson cross section
and $L=4\pi r^2\Gamma^2 n m_pc^3$ is the isotropic equivalent of
the kinetic luminosity of the jet (it approximately equals the 
total luminosity at $r>R_s$). The photosphere radius 
$\Rph\approx 10^{10}L_{52}\Gamma_3^{-3}$~cm 
is larger than $R_s$ for 
$\Gamma<10^3 L_{52}^{1/4}r_{0,7}^{-1/4}$. 
The radiation luminosity released at the photosphere 
is $L_\gamma\approx (\Rph/R_s)^{-2/3} L$,
and the mean energy of the escaping photons is given by
\be
\label{eq:Eph1}
  \Eph(\Rph)\sim \left(\frac{\Rph}{R_s}\right)^{-2/3}\Eph_0
   \approx 4\,\Gamma_3^{8/3} L_{52}^{-5/12} r_{0,7}^{1/6} {\rm ~MeV}.
\ee
The thermal radiation creates a bright burst with MeV peak if 
$\Gamma\sim 10^3$. The burst is weak for slower flows: 
$L_\gamma\propto\Gamma^{8/3}$.
%
\begin{figure}
\begin{center}
\includegraphics[width=3.3in]{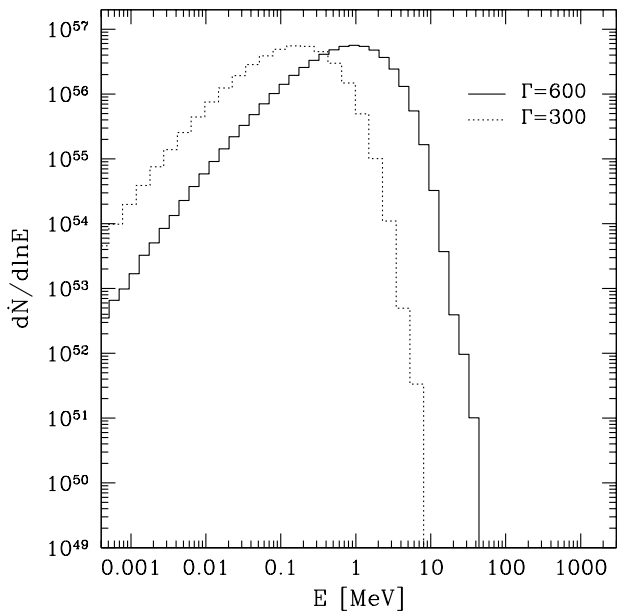}
\caption{Photon spectrum emitted by a passively cooling jet to distant 
observers. The jet has kinetic luminosity $L=10^{52}$~erg/s (isotropic 
equivalent) and initial size $r_0=10^7$~cm. 
It implies the total number flux of photons (isotropic equivalent)
$\dot{N}=1.6\times 10^{57}$~s$^{-1}$.
The photon spectrum 
has been calculated using the Monte-Carlo radiative transfer code. Two 
cases are shown: jets with asymptotic Lorentz factors $\Gamma=600$ and 300.
  The slopes of the spectra near 10~keV are close to 1.4, which corresponds 
  to photon index $\alpha=0.4$
The spectrum cuts off exponentially at $E>\Ep\sim 1$~MeV.
}
\end{center}
\end{figure}

We have calculated the radiation spectrum emerging from the passively
cooling jet using the Monte-Carlo code described in Appendix~B.
The scattering of the initial Planck spectrum was followed until the jet 
expanded to complete transparency. The spectrum of radiation received by 
distant observers is shown in Figure~1 for two example models.
In qualitative agreement with the above estimates, the jet with $\Gamma=600$
produces a bright GRB, whose spectrum peaks near MeV and cuts off 
exponentially. Two details are worth mentioning:

(i) Radiation emitted by passively cooling jets is not Planckian.
Observer sees different parts of the photosphere at different
angles, with different Doppler shifts. As a result, the low-energy slope 
of observed spectrum is softer than the Rayleigh-Jeans $\alpha=1$
(photon index). Isotropic emission in the jet frame would produce
$\alpha=0$, and the exact $\alpha$ is controlled by the photon angular
distribution near photosphere (e.g. $\alpha\approx 0.4$ near 10~keV in 
Fig.~1).
                                                                                
(ii) The standard description of adiabatic cooling predicts that 
radiation from a source at optical depth $\tT$ is cooled by the factor 
$A(\tT)=\tT^{-2/3}$ by the time the jet expands to its photospheric radius. 
The detailed transfer calculations give larger $A$. For example, $A(8)=0.58$ 
instead of $8^{-2/3}=0.25$ and $A(20)=0.39$ instead of $20^{-2/3}\approx 0.14$. 
The scaling $A\propto \tT^{-2/3}$ is maintained at large $\tT>10$.

The radiative transfer simulations illustrate and refine the standard 
fireball picture. They show that thermal radiation emitted by a passively 
cooling jet with $\Gamma>500$ peaks at $\Ep\sim 1$~MeV and carries away a 
significant fraction $\epsilon$ of the jet energy (e.g. $\epsilon\approx 1/4$ 
for the model with $\Gamma=600$ in Fig.~1). However, its spectrum cuts off 
exponentially at $E>\Ep$. Therefore, the model fails to explain the observed 
GRBs, whose spectra extend to energies $E\gg\Ep$.


\section{Neutron component and compound flows}

The picture described in \S~2 is incomplete, because it neglects the neutron 
component of the jet. In any plausible scenario of the GRB trigger, 
the central engine is dense, hot and neutron rich (Derishev, Kocharovsky
\& Kocharovsky 1999b; Beloborodov 2003). In particular, 
  accretion-disc models for GRBs predict a high neutron fraction
 (see Beloborodov 2008 for a review).
The neutron-rich matter is expected to enter the relativistic jet 
(although the details of this process may vary, see e.g. 
Levinson \& Eichler 2003; Metzger, Thompson \& Quataert 2008), 
and the neutron-proton jet initially accelerates as a single fluid 
where $n$ and $p$ are coupled by frequent nuclear collisions.
                                                                                
Neutrons and protons tend to combine into helium 
where the jet temperature drops to 140~keV. This process,
however, competes with rapid expansion and is marginally successful.
Collimation of the jet generally helps nucleosynthesis because it slows
down expansion (cf. Fig.~4 and 5 in Beloborodov 2003).
However, even in cases where helium production is complete,
some neutrons survive in jets with a neutron excess,
as helium production consumes equal numbers of $n$ and $p$.
In addition, free neutrons are produced by spallation 
of $\alpha$-particles at larger radii where the jet is heated.

The expansion of neutron-loaded jets generally leads to the formation 
of a {\it compound flow}: a slower neutron component with Lorentz factor 
$\Gamma_n$ is embedded in a faster proton flow with Lorentz factor 
$\Gamma>\Gamma_n$. 
The compound flow develops at the 
characteristic radius $R_n$ where the timescale for $n$-$p$ collisions 
becomes longer than the jet expansion time.

In particular, in jets 
that accelerate to $\Gamma>\Gcrit\approx 400L_{52}^{1/4}r_{0,7}^{-1/4}$ 
neutrons do not develop the full Lorentz factor --- their $\Gamma_n$ 
saturates at a smaller value (Derishev et al. 1999a; Fuller et al. 2000).
For example, a baryonic flow accelerated to $\Gamma=10^3$ can contain
neutrons with $\Gamma_n\sim 10^2$. In spite of the significant difference 
in Lorentz factors, the two components move together without any separation 
for a long time, because their velocities relative to the central engine 
are almost equal (the velocities practically equal $c$).

Even in jets with $\Gamma<\Gcrit$, compound flows with $\Gn<\Gamma$ 
are expected to form, because the jet is variable.
The neutron component does not participate in 
internal shocks that develop in variable jets. As a result, 
neutrons from the slow portions of the jet migrate across the shocks and 
penetrate the faster portions (M\'esz\'aros \& Rees 2000).
This internal mixing is caused by the short-timescale variability of the 
central engine that creates a non-uniform flow. The mixing occurs on scales
$\delta r\sim r/\Gamma^2$, much smaller than the total thickness of the 
ejected flow. Large variations of Lorentz factors\footnote{Large variations 
     on small scales are suggested by observed variability in GRBs.
     Large variations also generally help explain 
     the high efficiency of dissipation of internal motions in the jet 
     (e.g. Beloborodov 2000). 
}
produce a non-uniform compound flow with $\Gamma/\Gn\gg 1$.
%
\begin{figure}
\begin{center}
\includegraphics[width=3.2in]{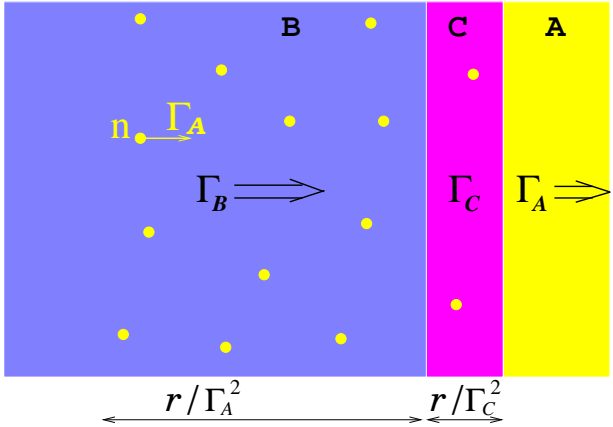}
\medskip
\caption{Faster flow $\B$ sweeps slower flow $\A$ and compresses it into
a shocked shell $\C$. Neutrons from flow $\A$ are not swept and instead 
penetrate flow $\B$. As a result a compound flow is formed: flow $\B$ 
contains a slower neutron component with $\Gn=\GA$. The penetration 
depth of neutrons is $\sim r/\GA^2$ in the lab frame; it is $(\GC/\GA)^2$ 
larger than the thickness of shocked region $\C$. 
}
\end{center}
\end{figure}

Neutron migration can be illustrated by the following simple model.
Suppose neutron-loaded flow $\A$ with Lorentz factor $\GA$
is followed by faster flow $\B$ with Lorentz factor $\GB\gg\GA$ (Fig.~2).
A shocked region $\C$ forms between the two flows;
the shocked plasma has Lorentz factor $\GC$ such that $\GA<\GC<\GB$.
Initially, the neutron component of flow $\A$ is coupled to protons by 
frequent collisions, so they behave as a single fluid.
When neutrons in flow $\A$ become collisionally decoupled, they are not 
swept into region $\C$ anymore. Instead, they penetrate region $\B$ with the 
relative Lorentz factor $\Grel=\frac{1}{2}(\GB/\GA+\GA/\GB)\approx\GB/2\GA$. 
The penetration/mixing length is $\sim(\GC/\GA)^2$ larger than the thickness 
of the shocked region $\C$.

Some of the penetrating neutrons collide with their new host flow.
Each collision dissipates the relative kinetic energy $(\Grel-1)m_pc^2$.
The number of collisions per baryon of flow $\B$ during the jet expansion 
timescale equals the collisional `optical depth' of the slow neutrons 
$\tau_n=n_n\signuc r/\Gn$, where $\sigma$ is the nuclear cross section.
At the beginning of neutron penetration $\tau_n\sim 1$ and a large heat 
is generated by collisions. The collisions decelerate flow $\B$ from $\GB$ 
to a new $\Gamma$, which is found from energy conservation in the static 
lab frame, $\tau_n\Gamma^2/2\GA\approx \GB$.\footnote{
    The decelerated flow with $\Gamma<\GB$ stores the heat of
    $\sim(\Gamma/2\GA)m_pc^2$ per baryon, and later tends to regain its 
    initial Lorentz factor $\GB$ as the heat converts back to bulk kinetic 
    energy via adiabatic cooling on the expansion timescale.}
This gives $\Gamma$ that is lower than the original $\GB$ by 
the factor $(\tau_n\GB/2\GA)^{-1/2}$ as long as $\tau_n>\Grel^{-1}$.

In summary, GRB jets are expected to contain a significant neutron component
(unless they are essentially baryon-free and completely dominated by
Poynting flux). At the radius $R_n$ where $\tau_n\sim 1$, collisions between 
neutrons and protons become rare and compound flows with $\Gamma>\Gn$ 
inevitably develop. The schematic picture in Figure~3 indicates 
the main characteristic radii of the jet.
The rare nuclear collisions in the region $\tau_n<1$
dissipate huge energy, comparable to the total energy of the jet. 
The dissipation efficiency of collisions is $(\Grel-1)\tau_n$.
It may exceed 100 per cent as the collisionally decelerated jet tends to 
regain its initial Lorentz factor via adiabatic cooling and re-dissipate 
its energy. Below we explore how collisional dissipation affects the 
jet radiation.

\begin{figure}
\begin{center}
\includegraphics[width=3.37in]{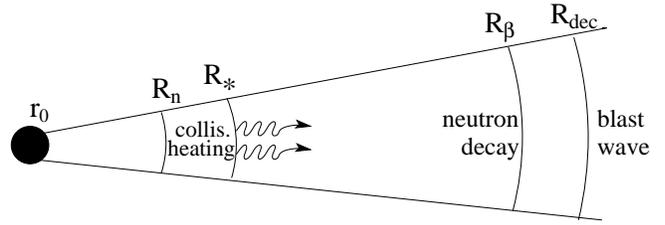}
\medskip
\caption{Schematic picture of a baryonic jet.
The jet starts to accelerate at radius $r_0$.
Compound flow with $\Gn<\Gamma$ forms at radius $R_n$ (eq.~\ref{eq:Rn})
and strong collisional heating begins at this radius. 
The jet becomes transparent to radiation at the photosphere 
$\Rph\sim 20R_n$ (eq.~\ref{eq:Rph}); its position is regulated by 
$e^\pm$ creation in the heated region.
The figure also shows the mean radius of neutron decay,
$R_\beta=3\times 10^{15}(\Gn/100)$~cm, and radius $R_{\rm dec}$ where
the jet starts to decelerate because of the interaction with an external
medium. The photospheric emission 
is released at $\Rph$, and its spectrum is strongly modified by 
sub-photospheric collisional heating. Collisional heating continues at 
$r>\Rph$, although with a smaller rate.
}
\end{center}
\end{figure}


\section{Radiative mechanism}

Hereafter we consider a simplified jet model: a neutron component with a 
single bulk Lorentz factor $\Gn$ is embedded in a fast proton component with 
constant Lorentz factor $\Gamma\gg\Gn$. The proper densities of the neutron 
and proton components will be denoted by $n_n$ and $n$, respectively.

\subsection{Inelastic nuclear collisions}

We consider collisions at radii where $\taunuc_n=n_n\signuc r/\Gn<1$, 
\begin{eqnarray}
\nonumber
   r>\Rdec & \equiv  & \frac{L_n\signuc}{4\pi m_pc^3\Gn^3} \\
           & \approx & \displaystyle{ 5\times 10^{11}\,
                    \left(\frac{L_n}{10^{52}\rm ~erg/s}\right)
                    \left(\frac{\Gn}{100}\right)^{-3} {\rm cm},}
\label{eq:Rn}
\end{eqnarray}
where $L_n=4\pi r^2 \Gn^2 n_nm_pc^2$ is the isotropic equivalent of the
kinetic luminosity of the neutron flow, and 
$\signuc\sim 3\times 10^{-26}$~cm$^2$ is the effective cross section 
for nuclear collisions.
The rate of collisions per unit volume (a Lorentz-invariant quantity) 
is given by 
\be
\label{eq:dn}
  \dn=nn_n\Grel\signuc c.
\ee 
Here $\Grel=\frac{1}{2}(\Gamma/\Gn+\Gn/\Gamma)\approx \Gamma/2\Gn$ is
the relative Lorentz factor of the neutron and proton components of the jet.

Collisions between neutrons and protons occur with significant $\Grel$ and 
hence have a large inelastic fraction $\fin\simgt 1/2$ (Amsler et al. 2008). 
The energy $\fin\Grel m_pc^2$ converts to mildly relativistic pions.
The data on $\pi^\pm$ multiplicity 
in $p$-$p$ collisions are found in e.g. Breakstone et al. (1984) and refs. 
therein; a similar multiplicity is expected for $n$-$p$ collisions. The 
total $\pi^\pm$ and $\pi^0$ multiplicity is larger by the factor of 3/2;
it is typically 5-6 for GRB jets.

The pions immediately 
decay: $\pi^\pm\rightarrow \mu^\pm+\nu_\mu \rightarrow e^\pm +\nu_e$ and 
$\pi^0\rightarrow \gamma+\gamma$. The produced neutrinos escape
with observed energies $\sim 0.1\Gamma$~GeV and carry away a fraction
$f_\nu\sim 1/2$ of the pion energy.\footnote{
    On average, neutrinos take $\sim 3/4$ of $\pi^\pm$ energy.
    The average fraction of $\pi^\pm$ and $\pi^0$ energy that is given to 
    neutrinos may be estimated as $\fnu\sim (2/3)(3/4)=1/2$.} 
This multi-GeV neutrino emission is an important prediction of the 
baryonic jet model (Derishev et al. 1999a; Bahcall \& M\'esz\'aros 2000; 
M\'esz\'aros \& Rees 2000), which may be verified or dismissed by future 
neutrino detectors. The existing upper limits from Super-Kamiokande 
experiment are $\sim 10$ times above the expected neutrino flux 
(Fukuda et al. 2002).
                                                                                
The fraction $1-f_\nu$ of pion energy is given to relativistic $e^\pm$ and 
high-energy $\gamma$-rays, which quickly convert to $e^\pm$ via 
$\gamma$-$\gamma$ reaction. Thus, the net result of one inelastic collision 
is the injection of several $e^\pm$ with a Lorentz factor 
$\gamma_0\sim m_\pi/m_e\approx 300$ in the rest frame of the plasma flow. 
The injected $e^\pm$ carry a significant fraction 
$f_\pm=\fin(1-\fnu)\approx 1/4$ of the collision energy $\Grel m_pc^2$.

Since neutrons move with a negative radial momentum in the frame of the 
proton flow, $e^\pm$ are injected with a negative  momentum. However, they 
become quasi-isotropic in the plasma frame after one Larmor rotation in 
the magnetic field of the jet (the field is transverse to the jet direction; 
its radial component is strongly suppressed as follows from magnetic flux 
conservation in the expanding flow). Any reasonable magnetic field advected 
from the central engine implies a very short gyration timescale 
$\gamma_0m_ec/eB$, and the net momentum of injected $e^\pm$ is immediately 
communicated to the plasma and vanishes in the plasma frame.

The energy of $e^\pm$ injected per unit volume per unit time 
equals $f_\pm\Grel m_pc^2\,\dn$. Practically all of this energy is quickly 
converted to radiation (see \S~4.2). As a result, radiation accumulates 
energy density with rate (measured in the plasma rest frame),
\be
   \dQnth=f_\pm\Grel m_pc^2\,\dn.
\ee
Here subscript `nth' stands for radiation produced by nonthermal $e^\pm$ 
that are generated by nuclear collisions. Using equation~(\ref{eq:dn}) and 
$\dQnth=c\Gamma\,dQ_{\rm nth}/dr$, one finds 
\be
\label{eq:dQnth}
    \frac{1}{nm_pc^2}\,\frac{d Q_{\rm nth}}{d\ln r}
   =\frac{f_\pm}{4}\,\frac{\Gamma}{\Gn}\,\tau_n
  \approx \frac{1}{16}\,\frac{\Gamma}{\Gn}\,\frac{\Rn}{r}.
\ee

\subsection{Radiative cooling of injected $e^\pm$}

The $e^\pm$ pairs injected by nuclear collisions have a large Lorentz factor
$\gamma_0\sim m_\pi/m_e$ and immediately radiate their energy via
Compton and/or synchrotron cooling.

\subsubsection{Compton cooling}

The timescale for Compton cooling of an electron with Lorentz factor 
$\gamma$ by radiation with energy density $\Ug$ is\footnote{This estimate 
    assumes Thomson scattering, i.e. neglects the Klein-Nishina correction 
    to the scattering cross section. The peak of GRB radiation is 
    $\Ep^\prime\sim {\rm MeV}/\Gamma\sim$~keV in the jet frame. Since 
    $\Ep^\prime\gamma<m_ec^2$ for all $\gamma\leq\gamma_0$, most of the 
    scattering by $e^\pm$ occurs in Thomson regime. Exact calculations of 
    radiative transfer with the full Klein-Nishina cross section are 
    performed in \S~5.} 
\be
\label{eq:tC}
   \tC=\frac{3m_ec}{4\Ug\sT \gamma}.
\ee
Radiation is initially present in GRB jets (\S~2), and $\Ug$ is further
increased as the radiation absorbs the energy of injected $e^\pm$.
Compton cooling timescale is shorter than the timescale of jet expansion
$\texp=r/c\Gamma$. Their ratio is
\be
   \frac{\tC}{\texp}=\frac{3}{4l\gamma},
\ee
where 
\be
\label{eq:l1}
   l\equiv\frac{\Ug}{m_ec^2}\,\sT\,\frac{r}{\Gamma}
\ee
is the dimensionless `compactness' parameter of the radiation field.
One can express $l$ as
 \be
 \label{eq:l2}
    l=\frac{m_p}{m_e}\,\eff\,\tp, \qquad \eff\equiv\frac{\Ug}{nm_pc^2}.
 \ee
Here $\eff$ is the fraction of the jet energy that is carried by radiation 
($\eff>0.1$ for the model proposed in this paper), and 
\be
\label{eq:tau_e1}
   \tp\equiv \frac{\sT n r}{\Gamma}
        =\frac{\sT}{\signuc}\,\frac{n}{n_n}\,\frac{\Gn}{\Gamma}\,\tau_n
        \approx \left(\frac{L}{5L_n}\right)
         \left(\frac{\Gamma}{5\Gn}\right)^{-3}\,\tau_n.
\ee
The compactness $l$ is high in the main heating region $\tau_n\simlt 1$,
and hence Compton cooling is fast, $\tC\ll\texp$.

The high compactness has another implication. Photons that are scattered 
by relativistic $e^\pm$ to energies $E^\prime\gg 1$~MeV in the jet frame
will not survive -- they will convert to secondary $e^\pm$ via reaction 
$\gamma+\gamma\rightarrow e^++e^-$. The development of $e^\pm$ cascade that 
accompanies Compton cooling of relativistic particles (Appendix~A) is 
described in detail by Svensson (1987) and Lightman \& Zdziarski (1987).
In this paper, the cascade is modeled numerically with our Monte-Carlo code. 
The typical multiplicity of secondary $e^\pm$ $\M_s$ is comparable to 60.  
The total multiplicity of $e^\pm$ created following one nuclear collision is
\be
  \M=\M_0\M_s\sim 10^2,
\ee
where
\be
 \M_0=\frac{f_\pm\Grel m_p}{\gamma_0m_e}\sim \frac{3}{4}\,\frac{\Gamma}{\Gn}
\ee
is the multiplicity of `primary' $e^\pm$ injected with Lorentz factor 
$\gamma_0\sim m_\pi/m_e$ following a nuclear collision.

\subsubsection{Synchrotron cooling}

In the presence of magnetic fields, the injected $e^\pm$ also experience 
synchrotron losses. The synchrotron cooling timescale is similar to 
equation~(\ref{eq:tC}) except that $\Ug$ in this equation is replaced by 
the magnetic energy density $U_B=B^2/8\pi$ (measured in the jet frame). 
The synchrotron losses dominate if $U_B>\Ug$. $U_B$ may be expressed as
\be
  U_B=\frac{B^2}{8\pi}=\frac{\epsilon_B L}{4\pi r^2\Gamma^2 c},
\ee
where $\epsilon_B$ is the magnetic fraction of the jet energy.
The typical energy of synchrotron photons in 
the plasma frame is $E_s^\prime=0.3\,\gamma_0^2\hbar eB/m_ec$. 
The corresponding energy in the lab frame, $E_s\approx \Gamma E_s^\prime$, 
is given by
\be 
\label{eq:E_s}
   E_s\approx 0.3\,\gamma_0^2 \frac{\hbar\,e}{m_ecr}\,
      \left(\frac{2\epsilon_B\,L}{c}\right)^{1/2}
     \approx 200\,r_{12}^{-1}\,\epsilon_B^{1/2}\,L_{52}^{1/2} {\rm ~keV},
\ee
where we substituted $\gamma_0\approx m_\pi/m_e\approx 300$. Synchrotron 
emission peaks in the region $r\simgt\Rn$ where heating peaks and most of 
$e^\pm$ are injected. Jets for which synchrotron cooling is significant 
(i.e. where it can compete with Compton cooling) have large $\epsilon_B$; 
then $E_s$ is comparable to the typical $\Ep$ of observed GRB spectra.
A similar $E_s\sim\Ep$ was found by Koers \& Giannios (2007). This feature 
of collisionally heated jets offers an additional mechanism for the 
preferential peak position at 0.1-1~MeV. 

Synchrotron emission from particles with low $\gamma\simlt 5$ is self-absorbed. 
These particles cannot be cooled by the synchrotron mechanism; they 
are Compton cooled.

\subsection{Optical depth of the jet}

In view of the strong $e^\pm$ loading and the large cross section for photon 
scattering $\sT\gg\signuc$, one may expect a large optical depth $\tT$ 
where the bulk of nuclear collisions occur. Note, however, that 
$\tT\propto\Gamma^{-3}$ while $\tau_n\propto\Gn^{-3}$. In compound flows 
$(\Gamma/\Gn)^{-3}\ll 1$; as a result, the collisionally heated jet with 
$\tau_n\simlt 1$ has a moderate $\tT$.

If no $e^\pm$ pairs were created, the optical depth of the compound flow 
would equal $\tp$ (eq.~\ref{eq:tau_e1}), which may be smaller than unity
at $r\simgt\Rn$. The actual optical depth is enhanced and dominated by 
$e^\pm$ created by the nonthermal cascade (Derishev et al. 1999a). 
The continually injected $e^\pm$
quickly cool down and accumulate at relatively low energies, forming a 
thermalized population that maintains a Maxwellian distribution via frequent 
Coulomb collisions between $e^\pm$.
It is convenient to express the rate of $e^\pm$ supply as
\be
\label{eq:dnpm}
  \dn_\pm=\M\dn=\frac{Y\,\dQnth}{m_ec^2},
\ee
where $\dQnth=f_\pm\Grel m_pc^2\dot{n}$ is the rate of energy injection in 
primary $e^\pm$, and $Y=\M_s/\gamma_0$ is the `pair yield' of the cascade. 
A minimum $Y_{\min}\sim \gamma_0^{-1}$ would be obtained when counting only 
the primary $e^\pm$ from pion decay. This may be appropriate for very 
strongly magnetized jets where synchrotron cooling of $e^\pm$ strongly 
dominates over Compton cooling ($U_B\gg\Ug$) and suppresses the $e^\pm$ 
cascade. In weakly magnetized jets with $U_B<\Ug$, the development of 
$e^\pm$ cascade gives $Y=\M_s/\gamma_0\sim 0.2$.

Let $n_\pm$ be the density of accumulated thermalized pairs.
Their annihilation rate is given by
\be
\label{eq:dnann}
  \dnann=\frac{3}{16}\,\sT cn_\pm^2.
\ee
This expression assumes $n_\pm>n$ and $kT<m_ec^2$; both assumptions are 
valid where annihilation is significant. The density of accumulated
$e^\pm$ evolves according to equation
\be
\label{eq:bal}
  \frac{\Gamma c}{r^2}\frac{d}{dr}\left(r^2n_\pm\right)
   = \dn_\pm-\dnann.
\ee
At the beginning of collisional heating, $\dn_\pm$ and $\dnann$ are both 
larger than the left side of equation~(\ref{eq:bal}), and the equilibrium
$\dn_\pm\approx\dnann$ is established, 
\be
  Yf_\pm\Grel^2\frac{m_p}{m_e}\signuc\,c\,n_n n = \frac{3}{16}\sT\,c\,n_\pm^2, 
\ee
which gives
\be
\label{eq:taupm}
   \tT(r)\equiv \frac{n_\pm \sT r}{\Gamma}
    =\left(\frac{4}{3}\frac{m_p}{m_e}\frac{\sT}{\signuc} 
        f_\pm Y\frac{L}{L_n}\right)^{1/2}\frac{\Gn}{\Gamma}\,\tau_n,
\ee
or, using $\tau_n=R_n/r$ and $f_\pm\approx 1/4$,
\be
\label{eq:tauT}
  \tT(r)=\tau_0\,\frac{R_n}{r}, \qquad 
  \tau_0\approx 20 \left(\frac{Y}{0.2}\right)^{1/2}
                   \left(\frac{L}{5L_n}\right)^{1/2}
                   \left(\frac{\Gamma}{5\Gn}\right)^{-1}.
\ee
$\tT$ stays near the equilibrium value $\propto\tau_n$ even after 
the annihilation timescale becomes long and the $e^\pm$ population 
freezes-out. This is the result of a coincidence: the annihilation 
equilibrium gives $\tT\propto r^{-1}$, which is also maintained 
when $\dn_\pm=\dnann=0$. Therefore, equation~(\ref{eq:tauT}) remains 
valid even at late stages when the jet becomes transparent to radiation.

The $e^\pm$ optical depth is maximum at the beginning (and peak) of 
collisional dissipation when $\tau_n\sim 1$. At this stage, $\Gamma/\Gn$
is limited by the deceleration effect of collisions (see the end of \S~3). 
In particular, the collision of flows $\A$ and $\B$ considered in \S~3 
gives a compound flow with $L\sim\LB$, $L_n\sim\LA$, 
and $\Gamma/\Gn\sim (\GB/\GA)^{1/2}$ at $r\simgt R_n$.
Then equation~(\ref{eq:tauT}) gives
\be
\label{eq:tauR}
    \tau_0\sim 20 \left(\frac{Y}{0.2}\right)^{1/2}
       \left(\frac{\GA}{\GB}\,\frac{\LB}{\LA}\right)^{1/2}.
\ee
It is reasonable to suppose $\LB/\LA>1$ when $\GB/\GA>1$ and expect 
$(\GB\LA/\GB\LB)^{1/2}\sim 1$ within a factor of a few. 

The result may be summarized by the simple approximate formula 
$\tT(x)\sim 20(Y/0.2)^{1/2}x^{-1}$, where $x=r/R_n$.
This estimate is a rather crude approximation (e.g. it neglects the
moderate adiabatic acceleration of the collisionally heated jet), 
yet it demonstrates an important feature: $\tT(x)$ weakly depends on
the parameters of the jet, as long as $\Gamma\gg\Gn$.
The estimate of $\tT$ also gives a simple expression for the photospheric 
radius,
\be
\label{eq:Rph}
    \Rph=\tau_0R_n\sim 20\left(\frac{Y}{0.2}\right)^{1/2} R_n.
\ee
 The radiation produced by collisional heating in the opaque region 
 $\Rn<r<\Rph$ is not buried by the optical depth. 
As demonstrated by the radiative transfer simulations in \S~5, it 
creates a powerful burst escaping to distant observers.

\subsection{Coulomb heating of thermalized $e^\pm$ by ions}

The thermalized $e^\pm$ population naturally tends to acquire the so-called 
Compton temperature in the radiation field, $\TC$, at which Compton cooling 
is balanced by Compton heating due to quantum recoil in scattering 
(e.g. Rybicky \& Lightman 1979). If no mechanism heats $e^\pm$, they would 
quickly reach Compton equilibrium with $kT_e=k\TC\sim 1$~keV.
This however does not happen, because 
  thermal $e^\pm$ are continually heated by
Coulomb collisions with protons. As a result, the $e^\pm$ temperature 
$T_e$ is maintained above $\TC$. Its value is calculated below; it 
satisfies $k\TC\ll kT_e\ll m_ec^2$ in the sub-photospheric heating region.

Nuclear collisions with $\Grel>1$ inevitably heat the proton component
of the jet to a relativistic temperature.\footnote{Nuclear collisions 
    also create a hot neutron component moving with the bulk Lorentz 
    factor $\Gamma$.}
The stirred protons acquire a non-Maxwellian distribution 
with a large fraction of protons having kinetic energies $\simgt m_pc^2$.
The temperature of the accumulated $e^\pm$ population 
is kept at a much smaller value by Compton cooling. In this `two-temperature' 
plasma, Coulomb collisions tend to transfer energy from protons to $e^\pm$. 
The thermal velocity of $e^\pm$ is well below $c$ and they may be
approximated as cold ($T_e=0$) when calculating the Coulomb energy losses
of the energetic protons. A proton with velocity $\bprot$ in the jet 
rest frame passes its energy to the cold $e^\pm$ background with rate 
(e.g. Ginzburg \& Syrovatskii 1964)
\be
\label{eq:dECoul}
   \dECoul=\frac{3}{2}\ln\Lambda\,\frac{\sT n_\pm m_ec^3}{\bprot},
\ee
where $\bprot\sim 1$, $\ln\Lambda=\ln(m_ec^2/\hbar\omega_p)\approx 20$ is 
the Coulomb logarithm, and $\omega_p=(4\pi n_\pm e^2/m_e)^{1/2}$. The net
rate of energy transfer from protons to the thermal $e^\pm$ plasma is 
\be
\label{eq:dQep1}
  \dQep\approx \frac{3}{2}\,\ln\Lambda\,n\sT n_\pm m_ec^3, 
\ee
which gives
\be
\label{eq:dQep}
   \frac{1}{nm_pc^2}\,\frac{d\Qep}{d\ln r}
        \approx\frac{3}{2}\ln\Lambda\,\frac{m_e}{m_p}\,\tT\approx 0.02\tT.
\ee
It is useful to compare $\dQep$ with $\dQnth$ (\S~4.1). From 
equations~(\ref{eq:dQnth}), (\ref{eq:dQep}), and (\ref{eq:tauT}) one finds,
\be
\label{eq:th_nth}
   \frac{\dQep}{\dQnth}\approx \left(\frac{L}{5L_n}\right)^{1/2}
                               \left(\frac{\Gamma}{5\Gn}\right)^{-2}
                               \left(\frac{Y}{0.1}\right)^{1/2}.
\ee  
The thermal and nonthermal heating rates are comparable.

The $e^\pm$ do not keep the heat $\Qep$ received from protons.
Instead, they immediately pass it to radiation via Compton scattering and 
remain at a temperature $kT_e\ll \Qep/n_\pm$. The value of $T_e$ is found 
from the balance between Coulomb heating and Compton cooling of $e^\pm$,
\be
   \frac{3}{2}\,n_\pm\frac{k(T_e-\TC)}{\tC}=\dQep,
\ee
where $\tC$ is given by equation~(\ref{eq:tC}) with $\gamma\approx 1$.
We have neglected the adiabatic cooling of $e^\pm$ because its rate is 
smaller than Coulomb heating and Compton cooling rates by the factor 
$\tC/\texp\ll 1$. Then we find
\be
\label{eq:Te}
  \Theta_e\equiv\frac{kT_e}{m_ec^2}
    \approx \frac{3m_e}{4m_p}\frac{\ln\Lambda}{\eff}+\frac{kT_{\rm C}}{m_ec^2}
    \approx \frac{0.01}{\eff}+\Theta_{\rm C},
\ee
where $\Theta_{\rm }=kT_{\rm C}/m_ec^2$ is the dimensionless Compton
temperature of the radiation field; $\Theta_{\rm C}\approx 0.007$ for the 
radiation spectrum calculated below (Fig.~5).

Kompaneets' $y$-parameter of thermal $e^\pm$ is given by 
\be
  y=4\tT(\Theta_e-\Theta_{\rm C})
  \approx \frac{0.04\tau_0}{\eff}\,\frac{R_n}{r}.
\ee
It is comparable to or below unity 
which shows that Compton cooling of $e^\pm$ occurs in the unsaturated 
regime. Thermal Comptonization has an important effect on the radiation 
spectrum, which is computed in \S~5.

\subsection{Distribution function of $e^\pm$}

\begin{figure}
\begin{center}
\includegraphics[width=3.3in]{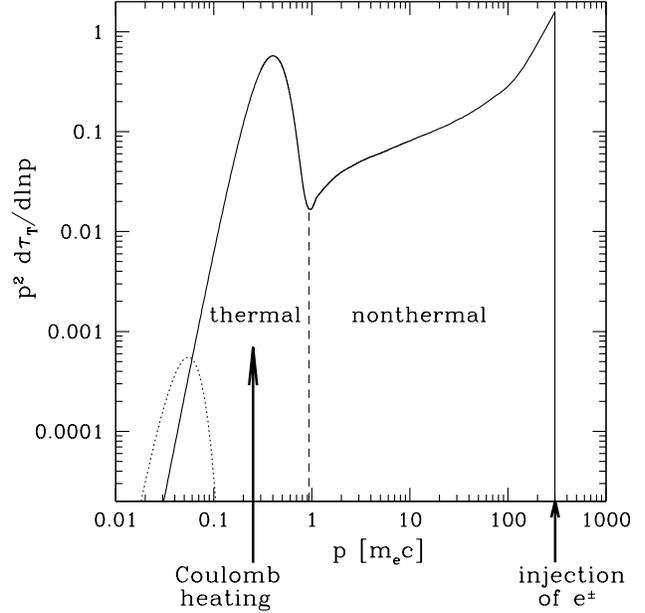}
\caption{
Momentum distribution of $e^\pm$.
{\it Dotted curve} shows the case of a passively cooling jet (the model
with $\Gamma=600$ from Fig.~1). {\it Solid curve} shows the case of
a collisionally heated jet at the same radius (see text). The distribution 
was calculated at $r=4R_n$.
{\it Vertical dashed line} indicates the boundary between the thermal and 
nonthermal parts of the distribution. The two parts make comparable 
contributions to the Compton amplification factor
$A\sim \int p^2\,(d\tT/dp)\, dp\sim 1$ that measures the average energy 
boost of photons in one scattering by the $e^\pm$ plasma.
The total/integrated optical depth $\tT$ at this radius is $\tT=5$;
it is strongly dominated by the thermal part of the distribution. 
}
\end{center}
\end{figure}

The local distribution function of $e^\pm$ is shaped by the processes of 
$e^\pm$ injection, cooling and thermalization, and Coulomb heating of 
thermalized $e^\pm$ by protons. The distribution is quasi-steady, i.e. 
it is established at a given radius on a timescale much shorter than the 
expansion timescale of the jet. It gradually changes as the jet expands. 
Figure~4 shows the momentum distribution of $e^\pm$ at radius $r=4R_n$ 
for a typical jet model. The jet has the same parameters as in Figure~1 
except that it now carries neutrons with $\Gn=100$ and 
$L_n=2\times 10^{51}$~erg/s. Synchrotron cooling was neglected in this 
example, i.e. the jet was assumed to be weakly magnetized, $\epsilon_B\ll 1$.
The distribution has been calculated by the Monte-Carlo code described 
in Appendix~B. The temperature of the thermal part $\Theta_e\approx 0.03$
is consistent with the analytical result (eq.~\ref{eq:Te}); it is 
self-regulated so that the balance is maintained between Coulomb heating 
and Compton cooling. The nonthermal part of the distribution is formed by 
the $e^\pm$ cascade that results from $e^\pm$ injection with 
$\gamma_0\approx m_\pi/m_e$. 

For comparison, the dotted curve in Figure~4 shows the electron 
distribution that is found at the same radius in the passively 
cooling neutron-free jet. The distribution is Maxwellian, and its 
temperature equals the temperature of the (Planckian) radiation field.

\subsection{Radiative efficiency of photospheric emission}

The evolution of radiation density $\Ug$ (measured in the plasma comoving 
frame) is given by equation 
\be
\label{eq:dUg}
  \frac{1}{r^2}\,\frac{d}{d\ln r}\left(r^2\Ug\right)
  =\left(\frac{d\Ug}{d\ln r}\right)_{\rm ad}+\frac{d\Qep}{d\ln r}
  +\frac{dQ_{\rm nth}}{d\ln r}.
\ee
The first term on the right side of this equation describes the adiabatic 
cooling of radiation; it equals $-(2/3)\Ug$ in the opaque zone and $0$ 
in the transparent zone (the exact behavior of this term near photosphere 
is obtained from the numerical simulation of radiative transfer). 
The second and third terms on the right side represent the energy 
received by $e^\pm$ plasma and converted to radiation.
Since practically all of the energy received by $e^\pm$ is passed 
to radiation, these terms effectively serve as sources of radiation energy.
Heating of the thermalized $e^\pm$ population by Coulomb collisions with 
protons $d\Qep/d\ln r$ is given by equation~(\ref{eq:dQep}).
Energy injection into the nonthermal $e^\pm$ tail $dQ_{\rm nth}/d\ln r$ 
is given by equation~(\ref{eq:dQnth}). Substituting these expressions
to equation~(\ref{eq:dUg}) we obtain the equation for $\eff\equiv\Ug/nm_pc^2$,
\be
\label{eq:eff_}
    x\,\frac{d\eff}{dx}=-q(x)\,\eff +\frac{\ath+\anth}{x},
\ee
where $\ath=0.02\tauR$ and $\anth=f_\pm\Grel/2$ are constants,
$x\equiv r/\Rn$, and $q(x)\equiv -(d\ln\Ug/d\ln r)_{\rm ad}$ is 
a dimensionless function that equals $2/3$ in the optically thick zone
$\tT\gg 1$ and approaches 0 at the photosphere. 
The quantity $\eff$ is the fraction of the jet energy carried by radiation;
it can also be written in the lab frame as 
\be
  \eff=\frac{L_\gamma}{L},
\ee 
where $L_\gamma=4\pi r^2\Gamma^2\Ug c$ is the isotropic equivalent of
radiation luminosity, and $L=4\pi r^2\Gamma^2 n m_pc^3$ is the isotropic
equivalent of the jet kinetic luminosity.

In the optically thick zone, where $q\approx 2/3$, equation~(\ref{eq:eff_}) 
can be solved analytically for $\eff(x)$,
\be
\label{eq:eff_a}
  \eff(x)=\frac{\eff_1+3a}{x^{2/3}}-\frac{3a}{x}, 
   \qquad 1<x<\frac{\Rph}{R_n},
\ee
where $\eff_1\equiv\eff|_{r=\Rn}$ and $a=\ath+\anth$. The solution may be 
used to estimate $\eff$ at the photosphere, $x_\star=\Rph/\Rn=\tauR$.
For jets with small $\eff_1$ one obtains
\be
\label{eq:eff_b}
   \effph
    \approx \left(0.06\,\tauR^{1/3}+\frac{0.2}{\tauR^{2/3}}\,\frac{\Gamma}{\Gn}
            \right)\left(1-\frac{1}{\tau_0^{1/3}}\right),
\ee
where $\tau_0\sim 20(Y/0.2)^{1/2}$ (Sect.~4.3).
Equation~(\ref{eq:eff_b}) estimates the net radiative 
efficiency of collisional heating in jets with $\Gamma\gg\Gn$,
taking into account the adiabatic cooling of radiation until the jet 
expands to transparency.\footnote{Equation~(\ref{eq:eff_b}) assumes the 
     adiabatic cooling $\propto r^{-2/3}$ at all $r<\Rph$. It overestimates
     the cooling effect -- the exact radiative transfer gives less cooling
     (see \S~2). Therefore, equation~(\ref{eq:eff_b}) underestimates $\effph$
     by a factor $\sim 2$.} 
For example, $\tau_0=20$ gives $\effph\approx 0.1+0.02\Gamma/\Gn$.
The two terms represent the contributions from the Coulomb heating of 
thermalized $e^\pm$ and the nonthermal $e^\pm$ injection.
For typical $\Gamma/\Gn\sim 3-10$ 
the total radiative efficiency $\effph=0.2-0.3$, with the thermal part 
comparable to the nonthermal part (cf. also eq.~\ref{eq:th_nth}).


\section{Radiation spectra from collisionally heated jets}

\subsection{Thermal and nonthermal Comptonization}

Suppose the jet cools passively at $r<\Rn$ and its thermal radiation
evolves as described in \S~2. The collisional heating begins at radius $\Rn$ 
(eq.~\ref{eq:Rn}) and quickly loads the jet with energetic $e^\pm$; their
typical distribution function is shown in Figure~4. Scattering of radiation 
by the injected $e^\pm$ dramatically changes the photon spectrum.

Consider a weakly magnetized jet with $U_B\ll \Ug$, when the synchrotron 
cooling of $e^\pm$ is negligible. Then the GRB spectrum forms via 
Comptonization of already existing thermal photons advected from the center 
of the explosion. Scattering conserves photon number and the average photon 
energy $\Eph$ in the lab frame can be expressed as (cf.~eqs.~\ref{eq:ratio} 
and~\ref{eq:Eph0})
\be
\label{eq:Eph}
  \Eph=\frac{\eff\,\Gamma n m_pc^2}{n_\gamma}=\eff\Eph_0
      \approx 4\,\eff\, r_{0,7}^{-1/2}L_{52}^{1/4} {\rm ~MeV},
\ee
where $\eff=L_\gamma/L$ is the fraction of jet energy carried by radiation.
The relation~(\ref{eq:Eph}) is common for all Comptonization models of GRBs
(e.g. Thompson 1994; Rees \& M\'esz\'aros 2005; Giannios \& Spruit 2007). 
It naturally explains the observed $\Eph\sim$~MeV, assuming a reasonable 
radiative efficiency $\eff\simgt 0.1$.
In the collisionally heated jet, $\Eph$ grows as photons receive energy 
via two branches of heating: thermal and nonthermal (\S~4). The corresponding 
heating rates {\it per photon} give
\be
\label{eq:HR_th}
   \left(\frac{d\Eph}{d\ln r}\right)_{\rm th}
          \approx 0.02\tau_0\Eph_0\frac{\Rn}{r}, 
\ee
\be
\label{eq:HR_nth}
   \left(\frac{d\Eph}{d\ln r}\right)_{\rm nth}
          \approx \frac{1}{16}\frac{\Gamma}{\Gn}\Eph_0\frac{\Rn}{r},
\ee
where $\Eph_0$ is given by equation~(\ref{eq:Eph0}).
The heating rates and the adiabatic cooling determine the evolution 
of $\Eph(r)$ in the collisionally heated jet.\footnote{Since 
        $\Eph/\Eph_0=\eff$ for a jet with a conserved photon number, 
        the equation for $\Eph(r)$ is immediately obtained from 
        eq.~(\ref{eq:eff_}).}
However, the known $\Eph$ does not yet determine the shape of the radiation 
spectrum. The spectrum depends on the details of Comptonization that need 
to be calculated.

The radiation spectra produced by Compton cooled $e^\pm$ cascades were 
previously studied in detail in the context of AGN accretion discs
(e.g. Svensson 1987; Lightman \& Zdziarski 1987; see also Appendix~A). 
The model was developed for static sources, and Comptonization of radiation 
in relativistic flows is different for two reasons. First, the optical depth 
evolves as the flow expands. Second, the GRB radiation moves 
together with the plasma and remains embedded in it until the jet reaches 
$r\sim\RD\sim\Gamma^2\Delta\sim 10^{16}$~cm (here
$\Delta/c\sim 1-10$~s is the typical duration of GRB jets). 
Collisional heating operates at smaller radii $r<\RD$, and the entire 
history of heating and Comptonization is `recorded' in the radiation field 
before it escapes the jet. The spectrum received by distant observers is 
the net result of multi-radius (multi-optical-depth) Comptonization in the 
expanding jet. In this respect, GRBs are similar to the relict radiation 
in the expanding universe.

The cooling rate of $e^\pm$ and their energy distribution at any given 
location depend on the local radiation field. Therefore, the evolution
of radiation and $e^\pm$ plasma must be calculated together. This is 
performed by the numerical code described in Appendix~B. The code is based 
on the Monte-Carlo method that solves the radiative transfer in a jet 
with self-consistent $e^\pm$ distribution function. For a given history 
of heating, the code calculates the evolution of $e^\pm$ and radiation in 
the expanding flow and finds the spectrum of photons escaping to distant 
observers.

As a typical example, consider the jet model from \S~2 with $\Gamma=600$, 
$L=10^{52}$~erg/s, and $r_0=10^7$~cm, but now let it contain a neutron 
component with $\Gn=100$ and $L_n=2\times 10^{51}$~erg/s. The collisional 
heating in this fiducial model begins at radius $\Rn\approx 10^{11}$~cm 
(eq.~\ref{eq:Rn}). 
Just before the onset of heating and $e^\pm$ creation, the passively cooling 
jet has $\Eph(\Rn)\approx 1$~MeV. The optical depth after the onset of 
collisional heating is $\tT(r)=(\Rn/r)\tau_0$ with $\tau_0\approx 20$ 
(eq.~\ref{eq:tauT}). The heating rates in equations~(\ref{eq:HR_th}) and 
(\ref{eq:HR_nth}) happen to be almost exactly equal:
$(d\Eph/d\ln r)_{\rm nth}\approx (d\Eph/d\ln r)_{\rm th}
\approx 1.5\,(\Rn/r)$~MeV.
%
\begin{figure}
\begin{center}
\includegraphics[width=3.3in]{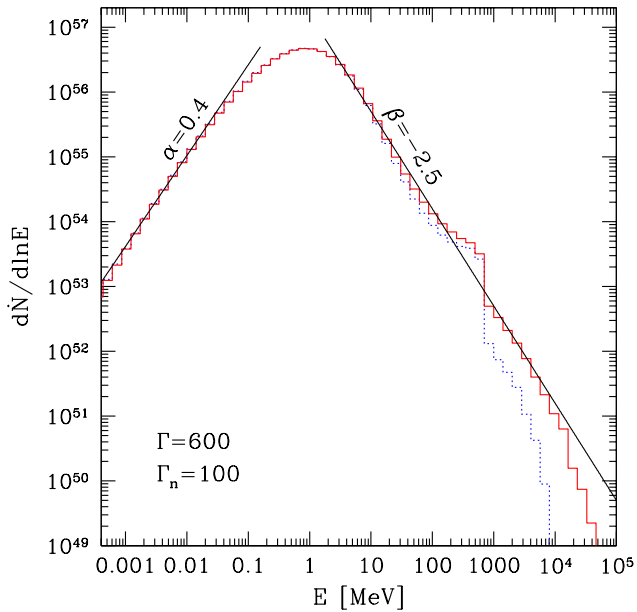}
\caption{Photon spectrum emitted by the collisionally heated jet 
({\it solid red histogram}).
The jet has $L=10^{52}$~erg/s, $r_0=10^7$~cm, $\Gamma=600$ 
(same as in Fig.~1), and carries neutrons with $\Gn=100$. 
{\it Black solid lines} indicate the slopes 
that correspond to photon indices $\alpha=0.4$ and $\beta=-2.5$.
A similar phenomenological spectrum 
is usually proposed to fit GRB observations (Band et al. 2009).
The feature near 0.5~GeV is the annihilation line.
  {\it Dotted blue histogram} shows the spectrum that would be produced 
  if nuclear collisions were `switched off' at $r>4R_n=0.2\Rph$, i.e.
  if $e^\pm$ injection was confined to radii $R_n<r<4R_n$. 
The figure does not take into account the cosmological redshift of the burst 
$z$; the redshifted spectrum will peak at $(1+z)^{-1}$~MeV instead of 1~MeV. 
}
\end{center}
\end{figure}

Figure~5 shows the spectrum of emitted radiation for the fiducial model.
Although it may not obvious from the figure, the Comptonized spectrum has 
two components, which correspond to the two parts of the $e^\pm$ distribution
function (cf. Fig.~4):

\medskip

\noindent
(i) Most photons are multiply scattered by the thermalized Coulomb-heated 
$e^\pm$ population and never scattered by the optically thin nonthermal tail.
This thermal Comptonization dominates the emitted spectrum at energies up 
to $2\Gamma kT_e\sim 20$~MeV and creates the spectrum slope 
$\beta\sim -(2.5-3)$. It corresponds to Kompaneets' parameter $y\sim 1$ 
that is regulated in the heated jet as discussed in \S~4.4.

\medskip

\noindent
(ii) A small fraction of photons are additionally scattered by the nonthermal
tail of $e^\pm$ distribution, which strongly boosts their energy. 
The nonthermal component dominates the radiation spectrum at high energies.
                                                                                
\medskip

A special feature of collisional heating is that the energy of the two
spectral components are comparable (eq.~\ref{eq:th_nth}). 
The nonthermal component smoothly extends the spectrum through 100~MeV to 
the GeV range. It is broad and additionally smoothed by partial 
downscattering in the optically thick plasma before the jet expands to 
transparency.

Most of collisional heating and Comptonization occurs where the jet is 
still opaque. 
The Comptonized radiation is released at the photosphere $\Rph$ and can be 
called `photospheric emission' (but see \S~5.3 below).
The average energy of escaping photons in the model shown in Figure~5
is $\Eph\approx 2$~MeV, which is half of $\Eph_0\approx 4$~MeV. 
This means that the net radiation efficiency of the burst is $\sim 50$\%,
i.e. the photospheric emission carries about half of the jet energy.\footnote{
    The energy given to photons by collisional heating in the region
    $\Rn<r<\Rph$ is $2\times 1.5$~MeV in the model shown in Fig.~5. 
    Together with the initial 1~MeV per photon at $\Rn$ this would make 
    $\Eph=4$~MeV, if there were no adiabatic cooling. Adiabatic cooling 
    at $r<\Rph$ reduces $\Eph$ by a factor of 2.} 

The ratio of the thermal and nonthermal Comptonization components in the 
observed spectrum is controlled by the parameter 
$\ww\approx 3\tau_0^{-1}\Gamma/\Gn$ 
(the ratio of eqs.~\ref{eq:HR_th} and \ref{eq:HR_nth}), 
where $\tau_0$ is likely to stay around 20 within a factor of a few
as long as the condition $\Gamma\gg\Gn$ is satisfied (\S~4.3). 
To investigate the sensitivity of the model predictions to expected 
variations in $\ww$, we calculated three models
with equal $\tau_0=20$ and different $\Gamma/\Gn=3$, 6 and 12.
They have $w\approx 0.5$, 1 and 2, correspondingly.
  We found similar spectra in all three cases, with slightly different 
  indices $\beta\sim 2.5\pm 0.2$
With increasing $w$, the nonthermal bump becomes more pronounced.
Large $\ww\gg 1$ are not, however, plausible (strong nonthermal heating 
is always accompanied by significant Coulomb heating in a realistic jet 
model).
Small $w$ are possible: $w$ can jump to zero if $\Gamma/\Gn$ decreases 
so that nuclear collisions become unable to produce pions. This case is 
discussed in \S~5.4 below.

\subsection{Annihilation line}

The annihilation reaction between thermalized $e^\pm$ produces photons
with energy $E^\prime\approx m_ec^2$ in the rest frame of the jet.
The number flux of annihilation photons (isotropic equivalent) 
in the lab frame is given by $\dNann=4\pi r^2c \Gamma\nann$, where
$\nann$ is the density of annihilation photons in the jet frame.
It obeys the equation,
\be
   \frac{d\dNann}{dr}=4\pi r^2\dnann.
\ee
Using equations~(\ref{eq:dnann}) and (\ref{eq:tauT}) one finds
\be
\label{eq:dNann_dx}
  \frac{d\dNann}{dx}=\frac{3\pi}{4}\,\frac{c\tau_0^2}{x^2}\,
                     \frac{\Gamma^2}{\sT}\,R_n,
\ee
where $x=r/R_n>1$ and $\tau_0$ is given by equations~(\ref{eq:taupm}), 
(\ref{eq:tauT}). Integrating equation~(\ref{eq:dNann_dx}) over $x$, 
one finds the net flux of annihilation photons emitted to infinity,
\be
  \dNann=\frac{f_\pm Y}{4\Gn}\frac{L}{m_ec^2}.
\ee
It is instructive to compare this result with the number flux of 
original thermal photons in the jet, $\dN$,
\be
  \frac{\dNann}{\dN}=\frac{f_\pm Y}{4\Gn}\,\frac{\Eph_0}{m_ec^2}
     \approx 2.5\times 10^{-4}\left(\frac{\Gn}{100}\right)^{-1}
     \left(\frac{Y}{0.2}\right)\left(\frac{\Eph_0}{\rm MeV}\right).
\ee
For our fiducial model shown in Figure~5, $\dNann/\dN\approx 10^{-3}$
creates a rather strong annihilation line that cuts off at 
$E=2\Gamma m_ec^2$.
Most of the annihilation photons are produced well below 
the photosphere. The resulting spectral feature has an extended red wing
due to Compton downscattering in the sub-photospheric region and the variation 
in the Doppler boost, which depends on the photon angle at the emission
(or last-scattering) point. 

In strongly magnetized jets, where synchrotron cooling
dominates over Compton cooling, the pair yield $Y$ is reduced (\S~4.3)
and the annihilation feature will be weak.

\subsection{$\gamma$-$\gamma$ opacity and emission at energies $E\gg$~GeV}

To a first approximation, one could neglect the heating at radii $r\gg R_n$,
and a similar spectrum would be obtained. For instance, suppose that 
nuclear collisions occur only in the region $R_n<r<4R_n$. The result
is shown by the dotted curve in Figure~5. The spectrum is significantly 
changed only at high energies: the number of photons above the threshold
for pair creation is suppressed.
This suppression is caused by the large compactness $l$ at small radii, 
which implies a large optical depth to $\gamma$-$\gamma$ absorption, 
$\tgg\gg 1$. 

The extension of the spectrum to $\sim 100$~GeV in the full model 
(solid curve) is due to the extension of nuclear collisions to large radii 
$r\sim 10^3R_n$, where $\tgg$ becomes small and high-energy photons are 
able to escape.\footnote{I thank Indrek Vurm for pointing out the effect 
    of continued collisional heating at large $r$ on the spectrum shape 
    in the GeV range.}
The smaller rate of nonthermal heating at large $r$ is compensated by the 
$\gamma$-$\gamma$ transparency at high energies. As a result, $\sim 10^{-3}$ 
of the jet energy is converted to escaping photons with energy comparable 
to 100~GeV.

A simple analytical estimate for the optical depth seen by a photon of 
energy $E$ at a radius $r$ is given by,
\be
   \tgg(E,r)=\frac{\sgg\,d\dN/d\ln E_t}{4\pi r c \Gamma^2}.
\ee
Here $\sgg\approx 10^{-25}$~cm$^2$ is the average cross-section for 
$\gamma$-$\gamma$ absorption by the target photons near the threshold,
$E_t/m_ec^2\sim 2\Gamma^2 (E/m_ec^2)^{-1}$. This estimate assumes a typical
angle $\theta\sim\Gamma^{-1}$ between the interacting photons. 
Approximating the spectrum of target photons by the Band function 
with $\Ep\sim 1$~MeV, one finds at $E_t>\Ep$
\be
  \frac{d\dN}{d\ln E_t}\approx 3\times 10^{57}
    \left(\frac{E_t}{m_ec^2}\right)^{1+\beta}\,L_{\gamma,52}\,{\rm ~s}^{-1},
\ee
where $\Lg$ is the total photon luminosity (isotropic equivalent). This gives,
\be
  \tgg(E,r)\approx \frac{2\times 10^3}{40^{-\beta-1}}\,r_{12}^{-1}
      L_{\gamma,52} \left(\frac{E}{10\rm ~GeV}\right)^{-\beta-1}
      \left(\frac{\Gamma}{600}\right)^{2\beta}.
\ee
For the typical $\beta\approx -2.5$, the radius of 
$\gamma$-$\gamma$ transparency (where $\tgg=1$) is given by
\be
   \Rgg(E)\sim 10^{13}\,\left(\frac{E}{10\rm ~GeV}\right)^{1.5}
     \left(\frac{\Gamma}{600}\right)^{-5}\,L_{\gamma,52}{\rm ~cm}.
\ee
The target photons absorbing 10-100-GeV photons have sub-GeV energy.
Most of them are emitted at $r\sim\Rph$ and hence the target radiation 
field at $r\gg\Rph$ is strongly collimated along the radial direction 
in the jet frame. This creates an ``escape cone'' for the high-energy 
radiation. The above estimates do not take into account this effect.
Detailed radiative-transfer calculations (as in Fig.~5) are needed
to get accurate results.

A crude estimate for the high-energy luminosity generated at large radii
may be obtained as follows. The rate of energy injection into the nonthermal
cascade is given by equation~(\ref{eq:dQnth}). This energy is deposited
to radiation via inverse Compton scattering.
At small radii, the injected nonthermal power is 
reprocessed by the cascade to smaller photon energies $E$ for which 
$\tgg(E)<1$. As a result, the escaping nonthermal luminosity at energies 
$\sim E$, $d\Lnth/d\ln E$, is roughly equal to the nonthermal power injected
at radii $\sim \Rgg(E)$,
\be
   \frac{1}{L}\,\frac{d\Lnth}{d\ln E}\sim \frac{1}{16}\,\frac{\Gamma}{\Gn}
          \,\frac{R_n}{\Rgg(E)}.
\ee
This gives
\be
\label{eq:Lnth}
  \frac{1}{L}\,\frac{d\Lnth}{d\ln E}\sim \frac{m_e}{16m_p}\,
   \frac{\sigma}{\sgg}\frac{L_n}{L}\frac{\Eph_0}{m_ec^2}
   \frac{\Gamma^{2-\beta}}{\Gn^4}
   \,\left(\frac{E}{2\Gamma m_ec^2}\right)^{1+\beta}.
\ee
This estimate
suggests that at high energies, where $\gamma$-$\gamma$ absorption is 
important, the photon spectrum $d\dot{N}/dE$ can steepen from the slope 
$\beta$ to slope $\beta-1$.

\subsection{Pure thermal Comptonization by Coulomb-heated electrons}

Some GRB jets may have the ratio $\Gamma/\Gn$ near unity.
Then nuclear collisions are not energetic enough to produce pions 
and the injection of relativistic $e^\pm$ pairs may not occur. 
The proton component can still be significantly heated by 
(elastic) $n$-$p$ collisions or other processes, e.g. internal shocks,
and the electron component is heated by Coulomb collisions with protons.
The electron heating rate $\dot{Q}_{\rm th}$ is proportional to the optical 
depth $\tT$ and can be significant below the photosphere. 
Radiation in such jets experiences pure thermal Comptonization, as
the scattering electrons have a Maxwellian distribution with temperature 
$T_e>\TC$ (\S~4.4).

Consider a jet with mildly 
relativistic protons; they heat the electrons according to 
equation~(\ref{eq:dQep}).\footnote{Protons with a smaller, sub-relativistic
   temperature pass faster their energy to electrons (Spitzer 1962). 
   Then, at large $\tT\gg 1$, {\it all} of the proton heat can be taken by 
   the electrons. In this case, the electron heating rate simply equals
   the dissipation rate in the jet. We consider here the case when 
   Coulomb collisions are slow enough to create a `bottleneck' for the 
   heat flow from protons to electrons to radiation. In this case, 
   electron heating is controlled by the rate of Coulomb collisions.}
The resulting radiation spectrum is shown in Figure~6. 
Our simulation assumed the heating rate per photon 
$(d\Eph/d\ln r)_{\rm th}=0.08\tT$~MeV and followed the evolution of 
radiation from radius $r=0.05\Rph$ where $\tT=20$.
The average photon energy $\Eph=1$~MeV was assumed at $r=0.05\Rph$.
It remained close to this value up to $r=\Rph$ where the jet became
transparent and released the Comptonized radiation. The emitted spectrum 
is suppressed exponentially above $E\simgt 2\Gamma kT_e\sim 20$~MeV where 
$kT_e\sim 15$~keV is the self-consistently calculated temperature in 
the main heating region.
The example shown in Figure~6 assumes $\Gamma=600$. We also ran a similar 
simulation for $\Gamma=300$; it gave a similar spectrum except that the 
cutoff occurred at $\sim 10$~MeV instead of 20~MeV.

\begin{figure}
\begin{center}
\includegraphics[width=3.3in]{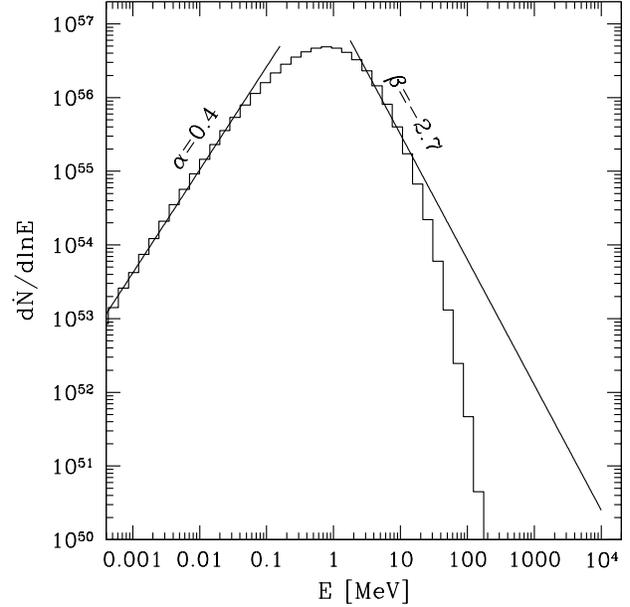}
\caption{Photon spectrum emitted by a jet with mildly relativistic 
internal motions, which do not lead to pion production. Protons have a 
mildly relativistic temperature in the sub-photospheric region and 
electrons are heated only by Coulomb collisions with protons (see text).
}
\end{center}
\end{figure}

Any mechanism that keeps protons hot in the sub-photospheric
region leads to similar Coulomb heating of electrons and a similar 
radiation spectrum. 
For example, protons may be heated by internal shocks in the jet.
Internal shocks
occur at radii $r\sim \Gamma_{\min}^2\delta r$, where $\delta r$ is the 
scale of fluctuations and $\Gamma_{\min}$ is the Lorentz factor of the 
slower parts of the jet. The scale $\delta r$ 
may be as small as $\sim 10^6$~cm (the size of the central engine) or 
perhaps even smaller. Then internal shocks begin at a radius 
$r\sim 10^{10}(\Gamma_{\min}/100)^2(\delta r/10^6 {\rm ~cm})$~cm
where the jet may have a large $\tT$ even without production of $e^\pm$
pairs (cf.~eq.~\ref{eq:tau_e}; $\Gamma$ is the Lorentz factor of the 
shocked part of the jet).

\subsection{Strongly cooled and then re-heated radiation}

The standard picture of a passively cooling jet (\S~2) predicts that
radiation experiences strong adiabatic cooling before reaching $\Rn$
if $\Rn\gg R_s$, which occurs for modest $\Gamma$. 
Thus, a regime is possible where radiation is strongly cooled
before collisional re-heating.

\begin{figure}
\begin{center}
\includegraphics[width=3.3in]{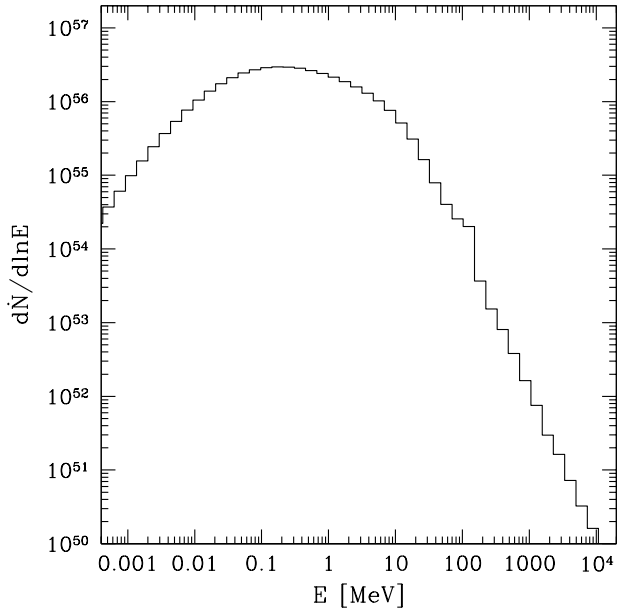}
\caption{Photon spectrum emitted by a jet whose thermal radiation was 
strongly cooled before collisional re-heating. The jet has the same $L$, 
$r_0$ and $\Gamma/\Gn$ as in Fig.~5, but its $\Gamma=150$ instead of 600.
}
\end{center}
\end{figure}

In this case, the emitted spectrum differs from Figure~5, as illustrated by 
the simulation shown in Figure~7.
It assumes that the jet has $\Gamma=150$, and all other parameters are
the same as in
Figure~5; in particular $\Gamma/\Gn=6$, which corresponds to $\Gn=25$.
The main difference caused by the smaller $\Gamma$ is that the radiation
temperature prior to the onset of heating drops to a low value
$kT_{\min}\approx 60$~eV (the adiabatically cooled $T_{\min}$ scales as 
$\Gamma^{8/3}$, see \S~2). It corresponds to $\Eph_{\min}\approx 0.025$~MeV 
and $\eff_{\min}\approx 6\times 10^{-3}$ at $r\simlt R_n$. 
The collisional re-heating at $r\simgt\Rn$ is still strong in the model, 
giving a significant radiative efficiency $\eff\approx 0.4$ and a significant 
$\Eph\approx 1.5$~MeV. 
  The resulting spectrum has a very broad peak above $\sim 0.1$~MeV. 
  It is shaped by thermal Comptonization with 
a large Compton amplification factor $A\sim 60$. This leads to a relatively 
hard slope $\beta\approx -1.4$ between $\Ep\sim 0.2$~MeV and 
$2\Gamma kT_e\sim 10$~MeV where $kT_e\sim 30$~keV is the self-regulated 
temperature of the $e^\pm$ plasma.

This simulation illustrates an interesting feature of Comptonization models
for GRB emission. Models with efficient re-heating do not give the simple 
Band-type spectrum with the MeV break if the mean photon energy $\Eph$ 
dropped much below MeV prior to re-heating. The cooling stage temporarily
creates an exponential break in the radiation spectrum at a low energy 
$E\ll 1$~MeV. Then re-heating and Comptonization increases $\Eph$ back to 
$\sim$~MeV; however, the recovery of $\Eph$ is achieved mainly by hardening 
the spectrum above the break, with only a minor shift in the break
position.\footnote{This is a robust result of unsaturated Comptonization
    in a relativistically expanding jet. Saturated Comptonization ($y\gg 1$) 
    would strongly shift the peak of the spectrum,
    however it appears to be not relevant to GRBs as it would require
    photon starvation while realistic GRB jets must advect a large number 
    of thermal photons from the central engine.}

We conclude that a long stage of adiabatic cooling at $r>R_s$ has a 
significant effect on the ultimate spectrum after re-heating. In reality, 
this effect may never occur if the jet does not passively cool between 
$R_s$ and $R_n$. Its proton component can be heated e.g. by internal shocks.
Then Coulomb-heated electrons will keep $\Eph$ and $\eff$ from falling,
as discussed in \S~5.4. Thus, it may be that even jets with 
$\Gamma\sim 100-300$ (modest by GRB standards) keep a significant 
$\eff\simgt 0.1$ prior to the onset of inelastic nuclear collisions 
at $R_n$. Then their emitted spectra will be similar to the spectra of 
high-$\Gamma$ jets shown in Figure~5.

\subsection{Impact of synchrotron emission and variability on the spectrum}

Two additional effects can change the observed spectra, in particular 
the slope $\alpha$ at $E<\Ep$:

\noindent
(i) A strong magnetic field adds synchrotron emission from $e^\pm$ pairs
injected by pion decay. It peaks at energies $E_s\simlt 1$~MeV 
(eq.~\ref{eq:E_s}) and can dominate at $E<\Ep$ because the synchrotron 
spectrum is relatively soft, $\alpha=-1/2$.

\noindent
(ii) Variable jets consist of many thin shells with different parameters,
and their radiation spectra vary on timescales as short as $10^{-4}$~s 
(in observer time). The superposition of many different instantaneous 
spectra is observed when the true instantaneous spectrum is not 
time-resolved. This tends to reduce the observed $\alpha$.

Models with synchrotron emission and variability will be published 
elsewhere.


\section{Discussion}

\subsection{Formation of GRB spectrum}

Formation of GRB spectrum is a long-standing problem. Much of the previous 
work focused on the optically thin internal-shock model 
(see Bosnjak, Daigne \& Dubus 2009 for recent detailed calculations).
The model posits that the observed $\gamma$-rays are nonthermal synchrotron 
emission from electrons accelerated at the shock fronts, and introduces 
phenomenological parameters of this process.
 Some key issues remained, however, unsettled. 
  Why is the nonthermal electron heating efficient?
Why do the reported spectra of GRBs usually peak near MeV?
Why are the low-energy slopes of some GRB spectra so hard ($\alpha\simgt 0$, 
significantly harder than e.g. synchrotron emission)?
As a possible solution, it was hypothesized that GRB spectra include a
bright photospheric component which results from strong sub-photospheric 
heating (e.g. Rees \& M\'esz\'aros 2005).

The results of the present paper suggest that the dominant component of 
GRB radiation comes from the photosphere.
Collisional heating naturally gives the photospheric emission a
Band-type spectrum (e.g. Fig.~5) without invoking unknown parameters 
apart from the Lorentz factors and the initial radius of the jet $r_0$. 
No fine-tuning of these parameters is required to produce the typical 
observed GRB spectrum. The radiative efficiency of collisional heating is 
large: it converts $\simgt 30$ per cent of the jet energy to escaping 
radiation.

In general, collisional heating depends on internal motions in the jet. 
One can imagine three possible regimes:

(A) The jet is steady (no fluctuations in Lorentz factor) and 
$\Gamma<\Gcrit$ (no neutron decoupling). Suppose also that there is
no magnetic dissipation. Then the outflowing plasma  
passively cools and emits the quasi-thermal spectrum shown in Figure~1.

(B) There are moderate fluctuations $\delta\Gamma/\Gamma<1$. 
Then the proton component of the jet is heated 
by internal shocks and nuclear collisions between protons and migrating 
neutrons. Electrons are heated by Coulomb collisions 
with protons, with a well defined rate (eq.~\ref{eq:dQep}).
The radiative efficiency of such jets is large if 
the protons are hot at radii $r\sim 0.1\Rph$ ---
then Coulomb collisions pass a large fraction of proton energy 
to electrons, and hence to radiation.
The radiation spectrum emitted by jets with Coulomb-heated electrons 
is shown in Figure~6. 

(C) If there are strong fluctuations $\delta\Gamma/\Gamma > 1$ or
the jet has a very high Lorentz factor $\Gamma>\Gcrit$, then compound 
flows form with $\Gamma/\Gn\gg 1$. Nuclear collisions in such jets create 
pions, which leads to injection of 
$e^\pm$ pairs with energies $\sim m_\pi c^2\approx 140$~MeV. 
The photospheric radius in this regime is regulated by the created $e^\pm$
pairs. The electron distribution function has an 
extended nonthermal tail, whose shape if determined by the radiative
cooling of $e^\pm$ (Fig.~4). The jet emits the radiation spectrum 
shown in Figure~5. The spectrum extends to very high energies with a slope 
$\beta\sim -2.5$ and has an annihilation line at $\sim \Gamma$~MeV whose 
amplitude is sensitive to $\Gn$.

Comparison of these theoretical expectations with available data suggests
that GRB jets are mainly in regime C (and may be in regime B in some bursts). 
The impact of collisional heating on the plasma and radiation components of 
the jet is significant in this regime.
The jet remains forever dominated by $e^\pm$, $n_\pm\sim 20 n$. 
The produced radiation remains embedded in the jet until 
it expands to $r\sim 10^{16}$~cm. Any additional heating processes occur 
in the radiation field already changed by the collisional heating.
                                                                                
The numerical simulations in this paper were performed for weakly magnetized 
jets, $\epsilon_B \ll 1$, where Compton cooling dominates over synchrotron 
cooling. Jets with large $\epsilon_B$ are expected to have 
  the same photospheric luminosity, 
with a similar spectrum that peaks near 1~MeV but has a smaller 
low-energy slope $\alpha$. 
Numerical models for strongly magnetized jets will be published elsewhere.

We showed numerical examples for a typical GRB jet with isotropic
equivalent of kinetic luminosity $L=10^{52}$~erg/s.
In our fiducial model we found $R_n\sim 10^{11}$~cm, $\Rph\sim 10^{12}$~cm
and $\Rgg(E)\sim 10^{13}(E/10{\rm~GeV})^{1.5}$~cm.
The model can be scaled to GRBs with different $L$.
Jets with fixed Lorentz factors (e.g. $\Gamma=600$ and $\Gn=100$) and 
fixed $L_n/L$ 
will have $\Rn\propto\Rph\propto\Rgg\propto L$, i.e. 
the characteristic radii will linearly scale with luminosity $L$.
The brightest observed GRBs have $L\sim 10^{54}$~erg/s, which 
leads to $R_n\sim 10^{13}$~cm, $\Rph\sim 10^{14}$~cm, and 
$\Rgg(E)\sim 10^{15}(E/10{\rm~GeV})^{1.5}$~cm. In spite of this big change,
the spectrum of produced radiation will be similar to that in Figure~5, 
because the {\it ratios} $\Rph/\Rn$ and $\Rgg/\Rn$ are important for the 
spectrum formation, not the values of radii.\footnote{ 
   The radius of collisional heating is limited by neutron decay 
   at $r\sim R_\beta=3\times 10^{15}(\Gn/100)$~cm, which 
   does not scale with $L$. For the brightest jets, the radius of 
   $\gamma$-$\gamma$ transparency at $10-100$~GeV becomes comparable to 
   $R_\beta$, which could affect the spectrum shape at the high-energy end.}
The value of $\Ep$ is likely to increase with $L$ (cf. eq.~\ref{eq:Eph}).

The slope $\alpha$ of the emitted spectrum is limited by the 
transfer effects discussed in \S~2. The hardest slope found in our
radiative transfer models near 10 keV is about $0.4$ (for comparison, 
a Planck spectrum would have $\alpha=1$). Practically all 
observed GRBs satisfy this limit (e.g. Preece et al. 2000). However, 
larger $\alpha$ were reported for a few bursts (Crider et al. 1997; 
Ghirlanda, Celotti \& Ghisellini 2003; Ryde et al. 2006). 
This suggests that in some bursts the jet may be inhomogeneous on tiny 
angular scales $\delta\theta<1/\Gamma$.

The relativistic jet is causally disconnected on scales 
$\delta r>r/\Gamma^2$, and different shells $\delta r$ can have different 
radiative history, with different $R_n$ and $\Rph$.
Observed GRB light-curves show strong variability in a broad range of
time-scales beginning from 0.1~ms.
The existence of very fast variability is naturally accommodated by our 
model. The shortest timescale of the photospheric emission is
$\delta t_{\rm var}\sim
\Gamma^{-2}(\Rph/c)\sim 10^{-4}(\Gamma/600)^{-2}R_{\star,12}$~s.

Note that observations of multi-GeV photons should not be used to constrain 
the radius of prompt emission $R_{\rm MeV}$ as done in Abdo et al. (2009a).  
They derive a minimum $\Gamma$ and a minimum $R_{\rm MeV}$ in GRB~080916C 
assuming that $R_{\rm MeV}$ is the same as $R_{\rm GeV}$ 
(for which $\gamma$-$\gamma$ transparency requires a large value).
In fact, MeV photons should not be assumed to come from the same radius 
as GeV photons,
even when the light-curves in the two bands are strongly correlated.
The general point is illustrated by the concrete model in the present paper.
The same plasma shell that emits MeV radiation at $\Rph$ can emit GeV 
photons when the shell expands to a larger radius $\Rgg$ (\S~5.3). 
The temporal correlation between MeV and GeV emission is preserved,
as photons emitted at different radii by the relativistically moving shell
arrive to observer almost simultaneously.
There is only a slight delay of the very-high-energy component emitted at 
$\Rgg$. This delay equals the observed time of the shell expansion from 
$\Rph$ to $\Rgg(E)$, which is $\Rgg/2\Gamma^2c\simlt 1$~s.

Similarly, observations of optical radiation that comes from a large radius 
$R_{\rm O}$ and is correlated with the prompt $\gamma$-rays 
(Racusin et al. 2008) cannot be used to constrain
the radius of the prompt $\gamma$-ray emission 
(see also the end of \S~6.3 below).

\subsection{Internal-shock heating}

The jet can be heated in the sub-photospheric region by multiple internal 
shocks as well as by nuclear collisions. A mildly relativistic shock front
heats protons to a mildly relativistic temperature and 
electrons to an ultra-relativistic temperature
if they receive a fraction $\epsilon_e\gg m_e/m_p$ of the dissipated 
energy. At the shock front, the electrons acquire 
a mean Lorentz factor $\ginj\sim \epsilon_e m_p/m_e$.
The details of this collisionless process are complicated
and can be studied by numerical simulations (e.g. Sironi \& Spitkovsky 2009).
The simulations suggest that shocks in a plasma with transverse
magnetic field $\epsilon_B>10^{-3}$ produce a rather narrow
(quasi-Maxwellian) distribution around $\ginj$.

The volume-averaged $e^\pm$ distribution function that results from shock 
heating is similar to that shown in Figure~4. 
The impulsive heating of electrons to $\gamma\sim\ginj$ is similar to the 
injection of $e^\pm$ by nuclear collisions, even though it is concentrated 
at the propagating shock front rather than distributed in volume. 
The heated electrons are 
quickly cooled behind the shock front and create an $e^\pm$ cascade.

As a result, the effect of internal-shock heating on radiation is in many 
respects similar to that of collisional heating, and the simulations in 
the present paper are useful for understanding this effect. 
If $\epsilon_e>10^{-2}$, $\ginj$ exceeds 20. In the sub-photospheric region, 
where compactness $l\gg 10$, the shape of $e^\pm$ distribution function 
at $\gamma>20$ is not important for Comptonization as the inverse-Compton 
emission from these electrons is anyway absorbed by the $\gamma$-$\gamma$ 
reaction. The shape of $e^\pm$ distribution 
is quite universal in weakly magnetized jets
-- it is controlled by the development of $e^\pm$ 
cascade in the same way for collisional heating and shock heating. 

The thermal part of $e^\pm$ distribution function must be nearly the
same in the two cases. It is regulated by the heat supply from protons
via Coulomb collisions and does not depend on what heats the protons --
internal shocks or nuclear collisions. Therefore, $\Theta_e$ is given 
by equation~(32) in either case.

The main difference between the nuclear collisional heating and shock 
heating is in the rate of electron energy injection. First note that the 
electron energy budget in shock heating is proportional to $\epsilon_e$. 
A small $\epsilon_e\ll 1$ implies that the normalization 
of the relativistic tail in the averaged $e^\pm$ distribution function is 
small compared to that produced by nuclear collisions, and its contribution 
to Comptonization is smaller.
Second, the dependence of shock heating on radius is uncertain as
it depends on the uncertain variability pattern of the central engine.
 Collisional dissipation has a special feature:
the `nonthermal' heating (injection of $e^\pm$ by nuclear collisions) 
$dQ_{\rm nth}/d\ln r\propto r^{-1}$ scales with $r$ in the same 
way as the `thermal' (Coulomb) heating $dQ_{\rm th}/d\ln r$ (\S~4), 
and their constant ratio $w$ is comparable to unity.
By contrast, the effective $w(r)$ for shock heating may vary,
leading to a different radiative-transfer solution for the
Comptonized spectrum.

Collisional heating alone gives a `minimal' emission model.
Shocks and magnetic dissipation are the usual
candidates for additional electron heating, 
which may create additional components of GRB radiation;
another mechanism for collisionless heating at large
radii is outlined in \S~6.3 below.
If the nonthermal electron population extends to $\gamma\gg 100$,
the relative contribution of synchrotron emission increases, as Compton losses
are suppressed by the Klein-Nishina reduction in scattering cross-section.
We did not simulate this situation here, and it can be done in the future.
The general setup of radiative-transfer calculations in this paper and 
the developed numerical code can be used for a broad 
class of emission models -- any combination of thermal 
heating and relativistic electron/positron injection 
in the expanding jet with any magnetization.

\subsection{Heating by neutron decay}

Neutrons carried by GRB jets eventually decay. Their large Lorentz 
factor implies a long decay time $\Gamma_n t_\beta$ where 
$t_\beta\approx 900$~s. The mean radius of $\beta$-decay is
\be
   R_\beta=ct_\beta\Gamma_n\approx 3\times 10^{15}
             \left(\frac{\Gamma_n}{100}\right) {\rm ~cm}.
\ee
The decay occurring inside the jet has a drag effect on the faster proton 
component and reduces its Lorentz factor (Rossi, Beloborodov \& Rees 2006). 
In essence, the decay injects relatively slow protons 
that are picked up by the jet with the relative Lorentz factor 
$\Grel=(1/2)(\Gamma/\Gn+\Gn/\Gamma)$. This can be described as inelastic
sharing of radial momentum between the fast jet and the decaying slow 
neutrons, which decelerates and heats the jet. The dissipation efficiency
of this process can exceed 100 per cent as the jet tends to use the heat to 
regain its Lorentz factor via adiabatic cooling and re-dissipate the energy. 

Most of neutrons decay near the radius $R_\beta$. However, a fraction 
$r/R_\beta$ decays at smaller radii $r<R_\beta$. Dissipation of $\sim 100$
per cent of the jet energy begins at radius $R_1\sim (\Gn/\Gamma)R_\beta$. 
Between $R_1$ and $R_\beta$, the jet decelerates in the background of 
decaying neutrons in a self-similar regime (resembling the deceleration of 
adiabatic blast waves), and its Lorentz factor decreases as $r^{-1/2}$. 
This strong dissipation may generate radiation. 

Note that the decaying neutrons create a perfect maser.
The new protons injected by $\beta$-decay appear in the plasma frame with 
momentum antiparallel to the flow direction and perpendicular to the 
magnetic field. They immediately begin to gyrate with Lorentz factor 
$\Grel$ and form a ring in the momentum space. This ring is a maser 
that amplifies low-frequency cyclotron waves in the plasma.
The maser instability develops on a short timescale proportional to
$\omega_B^{-1}$ where $\omega_B=eB/m_ec$ (e.g. Hoshino \& Arons 1991).
Damping of the excited waves heats the plasma. The waves may
also accelerate particles.
Extremely relativistic ion rings were previously studied near the
termination shocks of pulsar winds and proposed to accelerate leptons 
(Hoshino et al. 1992). 
A similar heating is observed in the interaction of comets with the solar 
wind. In this case, a compound flow is formed as the neutral gas around 
the comet penetrates the solar wind; ionization of the neutral particles 
effectively injects charges that move with a suprathermal velocity relative 
to the wind plasma and immediately begin Larmor rotation.

The $\beta$-decay and maser instability produce strong volume heating 
between $R_1$ and $R_\beta$. Coincidentally, at comparable radii, 
optical radiation can be released as self-absorption ceases in the optical 
band. Besides, the jet becomes transparent to very high-energy photons.
Thus, interesting radiative signatures may be expected. 
The radiative efficiency is, however, uncertain and likely smaller than 
the photospheric $\effph\sim 0.3-0.5$. The emission will occur in the 
optically thin regime and can be of the type modeled by Stern \& Poutanen 
(2004) and recently by Vurm \& Poutanen (2009). The study of possible 
radiative signatures of neutron decay between $R_1$ and $R_\beta$ is 
deferred to a future work. 

One feature of emission powered by neutron decay can be predicted.
The emission will arrive to distant observers with a slight delay with 
respect to the photospheric emission produced by the same neutrons via the 
collisional mechanism at $r\simlt\Rph$. As the jet expands from $\Rph$
to $r\gg \Rph$, a neutron with Lorentz factor $\Gn$ shifts with respect 
to the plasma jet a radial distance 
$\Delta r\approx (r-\Rph)(\Gn^{-2}-\Gamma^{-2})/2\approx r/2\Gn^2$,
which corresponds to observed delay
\begin{eqnarray}
\nonumber
   \Delta t_{\rm obs}\approx (1+z)\frac{r}{2\Gn^2c}
     \approx \frac{(1+z)t_\beta}{2\Gamma}\left(\frac{r}{R_1}\right)\\ 
     \approx \frac{(1+z)}{2}\left(\frac{\Gamma}{900}\right)^{-1}
               \left(\frac{r}{R_1}\right)  {\rm ~s}.
\end{eqnarray}
This neutron-drift delay appears to be consistent with the detected
delay $\Delta t_{\rm obs}\sim 2$~s of the prompt optical emission 
with respect to the main GRB pulses in the `naked-eye' GRB~080319B 
(Beskin et al. 2009). 

The delay of the GeV source detected by {\it Fermi} can have a 
similar, geometrical reason: if it operates at radii much larger 
than the source of MeV emission, its emission must be delayed.

\section*{Acknowledgments}

I grateful to I. Vurm, E. Derishev and the referee for 
comments that helped improve the manuscript.


\section*{References}

\small{

\noindent
Abdo A. A. et al., 2009a, Science, 323, 1688

\noindent
Abdo A. A. et al., 2009b, ApJ, 706, L138

\noindent
Amsler C. et al., Physics Lett. B667, 1 (2008)

\noindent
Asano K, Terasawa T., 2009, ApJ, 705, 1714

\noindent
Bahcall J. N., M\'esz\'aros P., 2000, Phys. Rev. Lett., 85, 1362
                                                                                
\noindent
Band D.~L. et al., 2009, ApJ, 701, 1673

\noindent
Beloborodov A. M., 2000, ApJ, 539, L25
                                                                                
\noindent
Beloborodov A. M., 2003, ApJ, 588, 931
                                                                                
\noindent
Beloborodov  A. M., 2008, AIP Conf. Proc., 1054, 51
                                                                                
\noindent
Beskin G. et al., 2009, submitted to Science (arXiv:0905.4431)

\noindent
Bosnjak Z., Daigne F., Dubus G. 2009, A\&A, in press 
 (arXiv:0811.2956)

\noindent
Breakstone A. et al., 1984, Phys. Rev. D, 30, 528

\noindent
Crider A. et al., 1997, ApJ, 479, L39

\noindent
Daigne F., Mochkovitch R., 2002, MNRAS, 336, 127

\noindent
Derishev E. V., Kocharovsky V. V., Kocharovsky Vl. V., 1999a, ApJ, 521, 640
                                                                                
\noindent
Derishev E. V., Kocharovsky V. V., Kocharovsky Vl. V., 1999b, A\&A, 345, L51
                                                                                
\noindent
Fukuda S. et al., 2002, ApJ, 578, 317
                                                                                
\noindent
Fuller G. M., Pruet J., Abazajian K., 2000, Phys. Rev. Lett., 85, 2673
                                                                                
\noindent
Ghisellini G., Celotti A., 1999, ApJ, 511, L93
                                                                                
\noindent
Ghirlanda G., Celotti A., Ghisellini G., 2003, A\&A, 406, 879

\noindent
Ghisellini G., Ghirlanda G., Nava L., Celotti A., 2009, MNRAS, in press
 (arXiv:0910.2459)

\noindent
Giannios D., Spruit H.~C., 2007, A\&A, 469, 1

\noindent
Hoshino M., Arons J., 1991, Phys. Fluids, B, 3, 818

\noindent
Hoshino M., Arons J., Gallant Y.~A., Langdon A.~B., 1992, ApJ, 390, 454
                                                                                
\noindent
Ioka K., Murase K., Toma K., Nagataki S., Nakamura T., 2007, 670, L77

\noindent
Koers H. B. J., Giannios D., 2007, A\&A, 471, 395

\noindent
Kumar P., Barniol Duran R., 2009, MNRAS, 400, L75

\noindent
Lemoine M., 2002, A\&A, 390, L31
                                                                                
\noindent
Levinson A., Eichler D., 2003, ApJ, 594, L19

\noindent
Lightman A. P., Zdziarski A., 1987, ApJ, 319, 643

\noindent
M\'esz\'aros P., Rees M. J., 2000, ApJ, 541, L5

\noindent
Metzger B.~D., Thompson T.~A., Quataert, E., 2008, ApJ, 676, 1130

\noindent
Paczy\'nski B., 1990, ApJ, 363, 218

\noindent
Pe'er A., M\'esz\'aros P., Rees M.~J., 2005, ApJ, 635, 476


\noindent
Preece R.~D. et al.,
2002, ApJS, 126, 19

\noindent
Pruet J., Woosley S. E., Hoffman R. D., 2003, ApJ, 586, 1254

\noindent
Racusin J. L. et al., 2008, Nature, 455, 183

\noindent
Rees M. J., M\'esz\'aros P., 2005, ApJ, 628, 847

\noindent
Rossi E.~M., Beloborodov A.~M., Rees M.~J., 2006, MNRAS, 369, 1797

\noindent
Ryde F., 2005, ApJ, 625, L95

\noindent
Ryde F. et al.,
2006, ApJ, 652, 1400 

\noindent
Ryde F. et al., 2010, ApJ, 709, L172

\noindent
Sironi L., Spitkovsky A., 2009, ApJ, 698, 1523

\noindent
Spitzer L., 1956, Physics of Fully Ionized Gases (New York: Interscience)
                                                                                
\noindent
Stern B.~E., Poutanen J., 2004, MNRAS, 352, 35
                                                                                
\noindent
Svensson R., 1987, MNRAS, 227, 403
                                                                                
\noindent
Thompson C., 1994, MNRAS, 270, 480
                                                                                
\noindent
Vurm I., Poutanen J., 2009, ApJ, 698, 293

}



\appendix
                                                                                
\section[]{Electron-positron cascade}

The compactness parameter $l$ is large in the sub-photospheric region 
$r<\Rph$ where most of collisional heating takes place (\S~4). 
It implies fast Compton cooling of $e^\pm$ and 
quick $\gamma$-$\gamma$ absorption of energetic photons.
Therefore, the timescale for the cascade development following the 
injection of an electron (or positron) with $\gamma_0\sim m_\pi/m_e$ is 
short compared with the jet expansion timescale. Then the distribution 
function of nonthermal $e^\pm$, $dn_\pm/d\gamma$, is locally 
(at a given $r$) qausi-steady and satisfies the equation
\be
\label{eq:fnth}
   \frac{d}{d\gamma}\left(\frac{dn_\pm}{d\gamma}\,\dot{\gamma}\right)
   =S(\gamma),
\ee
where $S(\gamma)=d\dn_\pm/d\gamma$ is the creation rate of secondary 
$e^\pm$, and $m_ec^2\dot{\gamma}(\gamma)$ is the energy loss rate of electron 
(or positron) with a Lorentz factor $\gamma$. The energy loss is due to 
Compton scattering, synchrotron emission, and Coulomb collisions with 
thermal electrons and positrons.\footnote{
     We neglect in this paper the possibility of energy exchange between 
     thermal plasma and nonthermal $e^\pm$ due to collective processes.}
Coulomb collisions dominate at small $\gamma\approx 1$ (and lead to
quick thermalization of non-relativistic $e^\pm$). In this Appendix, we 
focus on the relativistic tail of the $e^\pm$ distribution, where Coulomb 
collisions are negligible compared with Compton and synchrotron losses. Then,
\be 
  m_ec^2\dot{\gamma}\approx-\frac{4}{3}\,\sT c\,(\UKN+U_B)(\gamma^2-1),
\ee 
where $\UKN(\gamma)$ is the energy density of photons with energy 
$E^\prime\simlt m_ec^2/\gamma$, i.e. below the Klein-Nishina cutoff in 
the scattering cross section. 

The source function $S(\gamma)$ may be written as
\be
\label{eq:S}
  S(\gamma)=g(\gamma)\M_0\dot{n}, \qquad 
  \M_0=\frac{f_\pm\Grel m_p}{\gamma_0m_e}.
\ee
Here $\dot{n}$ is the rate of nuclear collisions in the compound flow
(eq.~\ref{eq:dn}), and $\M_0$ is the multiplicity of primary $e^\pm$ 
injected with Lorentz factor $\gamma_0$ following a nuclear collision. 
Then $g(\gamma)$ is a dimensionless function that represents the source 
of secondary $e^\pm$ created by one primary $e^-$ or $e^+$. This function 
is calculated numerically using Monte-Carlo simulations of the cascade.

Integration of equation~(\ref{eq:fnth}) yields 
\begin{eqnarray}
\label{eq:fnth1}
   \frac{d\tau_{\rm nth}}{d\gamma} \equiv
  \frac{\sT\,r}{\Gamma}\,\frac{dn_\pm}{d\gamma}
  =\frac{3}{8}\,\M_0\,\tau_n\,\frac{nm_ec^2}{\UKN+U_B}\,
  \frac{G(\gamma)}{\gamma^2-1}, \\
  G(\gamma)\equiv\int_{\gamma}^{\gamma_0}g(\gamma^\prime)\,d\gamma^\prime.
\end{eqnarray}
When $U_B\ll\Ug$, this equation becomes
\be
\label{eq:fnth2}
   \frac{\sT\,r}{\Gamma}\,\frac{dn_\pm}{d\gamma}
  =\frac{3}{8}\,\M_0\,\tau_n\,\frac{n}{n_\gamma}\,
  \frac{G(\gamma)}{\epKN(\gamma)(\gamma^2-1)},
\ee
where $n_\gamma/n\sim 10^5$ is the photon-to-baryon ratio 
(the main parameter of the GRB jet, see eq.~\ref{eq:ratio}), and
$\epKN(\gamma)\equiv\UKN/n_\gamma m_ec^2$ represents the mean dimensionless
energy per photon below the Klein-Nishina cutoff. The typical $\epKN$ in
the calculated models is near $3\times 10^{-3}$ and varies slowly 
with $\gamma$. The dimensionless function $G(\gamma)$ equals the number 
of secondary $e^\pm$ injected with Lorentz factor above a given $\gamma$
in the $e^\pm$ cascade triggered by one primary particle. In particular,
$G(1)=\M_s$ is the total number of secondary $e^\pm$. 
The function $G(\gamma)$ decreases from $G(1)=\M_s$ to $G(\gamma)\sim 1$ at 
$\gamma\simlt\gamma_0$, which implies a relatively slow dependence on 
$\gamma$, with the average slope 
$d\ln G/d\ln\gamma\approx -\ln\M_s/\ln\gamma_0\approx -0.7$. 
Then equation~(\ref{eq:fnth2}) implies that the optical depth of the 
nonthermal $e^\pm$ population $\tau_{\rm nth}$ sharply peaks at 
$\gamma\sim 1$. Its value is small, $\tau_{\rm nth}\ll 1$ (and much 
smaller than the optical depth of the thermalized $e^\pm$).
However, the effect of nonthermal population on radiation is
measured not by $\tau_{\rm nth}$, but by the Compton amplification factor 
$A=\int \gamma^2 d\tau_{\rm nth}$.
The amplification factor peaks at large $\gamma$,
\be
   \frac{dA}{d\ln\gamma}=\frac{3}{8}\,\M_0\,\tau_n\,
      \frac{n}{n_\gamma}\,
    \frac{\gamma\,G(\gamma)}{\epKN(\gamma)},
\ee
Particles with $\gamma^2\epKN\gg 1$ generate 
photons that are absorbed by the $\gamma-\gamma$ reaction. 
As a result, in the main heating region $r<\Rph$, particles with 
$\gamma\simgt 20$ contribute to the development of the $e^\pm$ 
cascade, while particles with $\gamma\simlt 20$
shape the scattered radiation spectrum.


\section[]{Numerical code}

The code is designed to simulate the self-consistent evolution of the 
radiation field and the $e^\pm$ plasma in the jet. For collisionally
heated jets considered in this paper, the following quantities are known 
at all radii: the injection rate of primary $e^\pm$ with 
$\gamma_0\sim 300$ (eq.~\ref{eq:dn}) and the corresponding energy injection 
rate (eq.~\ref{eq:dQnth}), the density of the accumulated thermalized 
$e^\pm$ component (eq.~\ref{eq:tauT}) and the heating rate of this 
component (eq.~\ref{eq:dQep}). The code aims to find the temperature of 
the thermalized $e^\pm$ population $T_e(r)$, the nonthermal tail of 
$e^\pm$ distribution, and the radiation field at all radii.

The calculation is split into two parts: 
(i) global radiative transfer in a jet with a given $e^\pm$ distribution
function, and (ii) calculation of $e^\pm$ distribution function 
for a given radiation field. The consistency between parts (i) and (ii) is 
reached via iterations as explained below. Note that part (i) is a global 
problem, while part (ii) is local and can be solved separately at all radii. 
Temperature $T_e$ and the nonthermal tail of $e^\pm$ distribution at a given 
$r$ are determined by the local radiation field, Coulomb heating rate, and 
$e^\pm$ injection rate. 

Radiation has a Planck spectrum at early stages of jet expansion
(i.e. at small radii), with the temperature determined by the initial size
of the jet and its energy. In the simulations, the initial thermal radiation 
is sampled by a large number of Planck photons ($\sim 10^9$), which are 
injected at a small radius and their scattering is followed until the photons 
escape.  The code can also simulate the injection of synchrotron photons
and follow them together with the Planck photons. In this paper,
only weakly magnetized simulations are presented, where Compton scattering 
is the dominant mechanism of spectrum formation, and synchrotron emission 
is neglected. Before the jet expands to transparency, the photons are 
multiply scattered and may be absorbed by the $\gamma$-$\gamma$ reaction. 
In each scattering event, the scattering electron is randomly drawn from 
the local $e^\pm$ distribution function, and the exact Compton cross section 
is used to randomly perform the scattering.

The radiative transfer is calculated in the static lab frame, assuming that 
the plasma flows in the radial direction with a bulk Lorentz factor $\Gamma$.
Since $\Gamma$ is large ($10^2-10^3$ in the simulations)
essentially all photons flow outward, and most of them have tiny angles
$\theta\sim\Gamma^{-1}$ with respect to the radial direction.
The radiation is essentially comoving with the plasma flow.
Therefore, one can view the transfer as the evolution of radiation in
time $t=r/c$ --- time and radius are almost equivalent choices for the
independent variable in the problem.
Between successive scatterings at radii $r_1$ and $r_2$, the photon
propagates along a straight line in the lab frame, and its angle with 
respect to the local direction of the radial jet, $\theta$, changes: 
$\sin\theta_2=(r_1/r_2)\,\sin\theta_1$.
This change automatically (and exactly) describes the adiabatic cooling of
radiation in the opaque zone.\footnote{
    This can be understood by considering the toy problem of coherent 
    and isotropic scattering in a cold jet.
    Then a scattering event does not change the photon energy in the 
    local plasma frame $E^\prime$; it only changes its angle.
    Between successive scatterings, the energy of the freely propagating 
    photon in the lab frame $E=const$, and 
    $E^\prime=E\Gamma(1-\beta\cos\theta)$ is decreasing because 
    of decreasing $\theta$. In addition, the propagating photon becomes 
    preferentially beamed outward in the plasma frame ($\theta^\prime$ 
    decreases). Next scattering again randomizes $\cos\theta^\prime$ and 
    destroys the preferential beaming, suddenly increasing (on average)
    $\theta^\prime$. As a result, the next scattering on average reduces the 
    photon energy $E=E^\prime\Gamma(1+\beta\cos\theta^\prime)$ 
    in the lab frame.}

When solving the radiative transfer with a trial $T_e(r)$, we find
the energy gained by radiation (per photon) from scattering on
thermal $e^\pm$.
This is done by defining a radial grid $r_i$ and accumulating statistics 
of scattering in each bin $\Delta\ln r$ during the Monte-Carlo simulation 
of the radiative transfer.
Thus, we evaluate $(d\Eph/d\ln r)_{\rm th}(r)$ for our trial model.
If it exceeds the required $(d\Eph/d\ln r)_{\rm th}$ (given in 
eq.~\ref{eq:HR_th}) we reduce $T_e(r)$ in the next iteration.

The nonthermal tail is given by equation~(\ref{eq:fnth}), which contains
the source function $S(\gamma)$ with shape $g(\gamma)$ (eq.~\ref{eq:S}). 
We find $g(\gamma)$ numerically using 
the Monte-Carlo simulation of the cascade in the local radiation field 
(which is known after calculating the radiative transfer in the 
previous iteration). The distribution of nonthermal $e^\pm$ at
small $\gamma$ is affected by Coulomb collisions with the thermalized 
$e^\pm$ population. This effect is included by adding the Coulomb losses 
to $\dot{\gamma}$ in equation~(\ref{eq:fnth}). The losses are evaluated
approximately by assuming a cold $e^\pm$ background;\footnote{
    The exact shape of the $e^\pm$ distribution function in the region  
    connecting the thermal and nonthermal parts requires the full treatment
    of Coulomb collisions with a finite-temperature plasma. However, 
    this region radiates very little, and the approximate matching 
    of the thermal and nonthermal components (as in Fig.~4) is 
    sufficient for the radiative transfer simulations.}
they are given by equation~(\ref{eq:dECoul}) (Coulomb losses are similar for 
energetic protons and $e^\pm$).  
The updated nonthermal tail and $T_e(r)$ are used in 
the calculation of radiative transfer in the next iteration.
5-10 iterations are usually sufficient to accurately find the self-consistent
solution for the radiative transfer and the $e^\pm$ distribution function.

The iterative method also allows the code to achieve a self-consistent 
treatment of $\gamma$-$\gamma$ absorption, which is a nonlinear effect.
When the code solves radiative transfer, the opacity to $\gamma$-$\gamma$ 
absorption $\kgg$ is evaluated using the radiation field saved from the 
previous iteration (or an initial guess, for the first trial). 
In each radial bin $\Delta \ln r$, the radiation field 
is saved on a grid in the ($\theta,E$)-space as a collection of 
$n_\theta\times n_E$ `monocromatic beams'. 
The $\gamma$-$\gamma$ opacity seen by a 
given photon with energy $E_0$ propagating at angle $\theta_0$ is calculated
by integrating over all target `beams' $(\Delta\theta,\Delta E)$ 
(with a random azimuthal angle) that are above the threshold for reaction 
$\gamma+\gamma\rightarrow e^++e^-$.
As the photon $(E_0,\theta_0)$ traverses the radial bin $\Delta\ln r$,
the probability of its survival is $\exp(-\kgg\Delta r/\cos\theta_0)$. 

\end{document}